\documentclass[onecolumn,sort&compress,numbers]{els-mrw-mod}
\usepackage{amsmath,amssymb,amsfonts,amsthm,makeidx,graphicx}
\usepackage{txfonts}
\usepackage{helvet}
\usepackage{bbold}
\usepackage[usenames,dvipsnames,svgnames,table]{xcolor}
\usepackage{slashed}
\usepackage{tikz}
\usetikzlibrary{positioning}
\usepackage{hyperref}

\newcommand{\de}{\delta} 

\newcommand{\ba}{\begin{align}}
\newcommand{\ea}{\end{align}}	
\newcommand{\eref}[1]{Eq.~(\ref{#1})}
\newcommand{\fref}[1]{Fig.~\ref{#1}}
\newcommand{\tref}[1]{Tab.~\ref{#1}}
\newcommand{\nnnl}{\nonumber\\}	

\begin{document}

\chapter{A beginner's guide to functional methods in particle physics}\label{chap1}

\author[1]{Markus Q. Huber}

\address[1]{\orgname{Justus-Liebig-Universit\"at Giessen}, \orgdiv{Institut f\"ur Theoretische Physik}, 
\orgaddress{Heinrich-Buff-Ring 16, 35392 Giessen, Germany}}

\maketitle

\begin{abstract}[Abstract]
	Functional methods like Dyson-Schwinger equations, the $n$PI effective action formalism, bound state equations and the functional renormalization group are versatile tools to study quantum field theories.
	They are exact, nonperturbative equations but have to be truncated for practical calculations.
	After a general introduction, I focus on their use in particle physics and discuss common truncations and solution techniques.
	The complete process from choosing a truncation to calculating observable quantities is exemplified by means of the glueball spectrum.
\end{abstract}

\begin{keywords}
 	functional methods\sep Dyson-Schwinger equations \sep $n$PI effective action\sep bound state equations\sep functional renormalization group
\end{keywords}

\begin{glossary}[Nomenclature]
	\begin{tabular}{@{}lp{34pc}@{}}
		BSE & Bethe-Salpeter equation\\
		DSE & Dyson-Schwinger equation\\
		EOM & Equation of motion\\
		FRG & Functional renormalization group\\
		IR & Infrared\\
		$n$PI & $n$-particle irreducible\\
		QCD & Quantum chromodynamics\\
		QED & Quantum electrodynamics\\
		QFT & Quantum field theory\\
		RG & Renormalization group\\
		STI & Slavnov-Taylor identity\\
		UV & Ultraviolet\\
	\end{tabular}
\end{glossary}

\section*{Objectives}
\begin{itemize}
	\item Understand the origin and meaning of different functional equations.
	\item How to use functional equations: overview of truncation schemes and solution techniques.
	\item How to use equations of motion and bound state equations to calculate a mass spectrum.
\end{itemize}


\section{Introduction}
\label{sec:intro}

Functional methods encompass several systems of equations in quantum field theories.
Their nonperturbative nature makes them particularly useful for studying situations where perturbative methods fail.
Thus, they are used in many areas of physics, among them particle physics, condensed matter physics, and quantum gravity.
One strength is their ability to provide insights into the underlying mechanisms of physical systems, while they also allow for quantitative calculations.

In this article I provide an overview over the different types of functional methods used in modern particle physics research and how they are applied in practice.
Functional equations are exact relations between correlation functions which are usually formulated in momentum space and often involve loop integrals.
Analytic solutions of these equations exist only in limiting cases, like the UV or IR limit, or under severe approximations.
For quantitative results one has to resort to numerical methods.
Depending on the problem, the necessary computational resources can range up to high-performance clusters, but many problems are actually far less demanding.
Nevertheless, the increase in available computational power in the last decades was surely beneficial for progress in the field.

Here we will focus on the following variants of functional equations.
Equations of motion of correlation functions are integral equations that relate different correlation functions to each other.
Functional renormalization group equations are integro-differential equations in which the introduction of an artificial scale allows integrating out quantum fluctuations successively from high to low energies.
Bound state equations are also integral equations and describe how constituents form composite objects like mesons or baryons.
These different types of functional equations each require dedicated methods to solve them.
Before such methods can be put to work, the equations have to be truncated, viz., the system of equations, which is infinitely large, has to be brought into a manageable, finite form.
This step constitutes one of the main challenges when working with functional methods and requires physical insight and a good understanding of the underlying theory.
How to truncate the equations and which solution methods exist will be discussed as well.

A main advantage of functional methods is that they are formulated in the continuum.
This avoids the complications arising from discretizing space-time like finite volume effects, the breaking of rotational symmetry, or the appearance of fermion doublers.
The challenges of functional methods come from a different direction, namely the aforementioned necessity to truncate the infinitely large systems of equations.
One possibility to tackle that consists in the use of phenomenologically motivated ans\"atze for certain correlation functions to decouple the equations and obtain a finite system.
This is also called bottom-up approach.
While being successfully applied for many problems, the disadvantage is that the employed ans\"atze may deviate from the true forms of the correlation functions they represent.
Hence approximation schemes are typically developed for specific applications.
Another approach is to calculate more correlation functions and reduce as much as possible or even eliminate the dependence on models, aiming at so-called apparent convergence.
Such a top-down approach is technically more involved due to the larger systems of equations and explicit calculations of correlation functions instead of using models.
The number of correlation functions which need to be included dynamically can only be estimated in advance and one needs to check the validity of a truncation.
If a calculation does not require any external input, it is called self-contained.
Between these two extreme variants there is a range of possibilities to combine both approaches or include results of correlation functions from other sources.

The main part of this article is organized in three sections.
Depending on the reader's background and interest, different sections are recommended to start with.
\begin{itemize}
\item Sec.~\ref{sec:funcEqs} gives an overview over functional equations and their theoretical background.
This section is mainly aimed at readers completely unfamiliar or only somewhat familiar with such methods.
The derivation of equations of motion, bound state equations, and functional renormalization group equations is sketched and some other functional methods are shortly mentioned.
\item In Sec.~\ref{sec:workingWithFuncEqs} we will see how to get from the formal and exact equations to actual solutions thereof.
It discusses several truncations schemes and solution techniques.
Readers looking for a guide on how to work with functional equations in practice will find information there.
\item A complete working example is provided in Sec.~\ref{sec:glueballSpectrum}.
It discusses the choice of equations and their truncations and how to solve them numerically to obtain the glueball spectrum of pure Yang-Mills theory.
This section is recommended for readers who are just interested in a bird's eye view of how functional equations are put to work using a concrete example.
\end{itemize}
Finally, Sec.~\ref{sec:conclusions} summarizes the main points and lists some applications of functional equations in other areas of physics.

\begin{BoxTypeA}[box:furtherReading]{Further reading}

Applications of functional equations in particle physics are vast and this article can only provide a glimpse at them.
In the following, some entry points for further reading are collected without any claim of completeness and focusing on the last 25 years.

Anyone who is interested in more details on the derivation of functional equations including examples and review articles is referred to \cite{Alkofer:2000wg,Maris:2003vk,Fischer:2006ub,Alkofer:2008nt,Swanson:2010pw,Maas:2011se,Huber:2011qr,Bashir:2012fs,Huber:2018ned,Huber:2019dkb,Ferreira:2025anh} for Dyson-Schwinger equations, \cite{Cornwall:1974vz,Berges:2004pu,Carrington:2010qq,Huber:2018ned} for $n$PI effective actions, \cite{Alkofer:2000wg,Maris:2003vk,Sanchis-Alepuz:2015tha,Bashir:2012fs,Eichmann:2016yit,Eichmann:2020oqt,Eichmann:2025wgs} for bound state equations, and \cite{Aoki:2000wm,Berges:2000ew,Polonyi:2001se,Pawlowski:2005xe,Gies:2006wv,Schaefer:2006sr,Delamotte:2007pf,Rosten:2010vm,Kopietz:2010zz,Braun:2011pp,Huber:2011qr,Huber:2019dkb,Dupuis:2020fhh} for the functional renormalization group.
For applications of functional equations at nonzero temperature and density, see, e.g.,~\cite{Fischer:2018sdj,Dupuis:2020fhh,Fischer:2026uni}.

General texts on quantum field theory are, for example, \cite{Amit:1984ft,Rivers:1988pi,Peskin:1995}.
An overview over QCD and other strongly coupled theories can be found in \cite{Brambilla:2014jmp}.

\end{BoxTypeA}

\section{Functional equations}
\label{sec:funcEqs}

In this section various functional equations will be introduced.
First, the underlying quantum field theoretic background will be discussed.
This is largely textbook knowledge, see, for example, \cite{Amit:1984ft,Rivers:1988pi,Peskin:1995}, condensed to what is relevant for the following sections.
However, it also serves to fix the notation and to provide, for the reader unfamiliar with quantum field theory, an account of the concepts used in the following.
Some topics going beyond typical textbooks including $n$PI effective actions and effective average actions will be introduced in the subsequent sections.
Different sets of equations, namely equations of motion of correlation functions, bound state equations, and functional renormalization group equations will be discussed in turn, followed by a short overview of other functional methods.
Depending on the knowledge of the reader, the corresponding subsections may be skipped.

\subsection{Quantum field theory background}
\label{sec:QFT}

At the heart of a QFT is the Lagrangian density.
It encodes the dynamics of the fields in form of their propagators, interactions, and underlying symmetries.
For the sake of conciseness, we will work for now with a scalar field theory and ignore complications of fields with higher spin or more components.
The underlying structure of the equations, as becomes most manifest when represented in diagrammatic form, will become clear even in such a simple model.
Alternatively, one may also consider the scalar field $\phi$ as a superfield whose components are made up of any desired specific fields.
As it is most convenient, we will work with the path integral formulation and, like the majority of the literature, we will use Euclidean spacetime.

In the subsequent subsection, we will consider the Lagrangian density of a scalar QFT with cubic and quartic interactions.
Its Lagrangian density reads:
\begin{align}\label{eq:scalarLagrangian}
	\mathcal{L}(\phi(x), \partial_\mu\phi(x)) &= \frac{1}{2}\phi\left(-\partial^2 + m^2\right)\phi - \frac{g_3}{3!}\phi^3 - \frac{g_4}{4!}\phi^4.
\end{align}
The action is then given by
\begin{align}\label{eq:scalarAction1}
	S[\phi] & = \int d^4x\, \mathcal{L}(\phi(x), \partial_\mu\phi(x)) = \frac{1}{2} \int_p \phi(p)(p^2 + m^2)\phi(-p) - \frac{g_3}{3!}\int_{pq}\phi(p)\phi(q)\phi(-p-q) - \frac{g_4}{4!}\int_{pqr}\phi(p)\phi(q)\phi(r)\phi(-p-q-r)\\\label{eq:scalarAction2}
	&  = \frac{1}{2!}S_{rs}\phi_r \phi_s - \frac{1}{3!}S_{rst}\phi_r \phi_s \phi_t - \frac{1}{4!}S_{rstu} \phi_r \phi_s \phi_t \phi_u,
\end{align}
where we introduced the coefficients $S_{rs}$, $S_{rst}$, and $S_{rstu}$ denoting the bare two-, three- and four-point functions and used a shorthand notation where summation of indices represents integration over position or, as in the form written here, momentum space.
This notation can later be generalized straightforwardly to indices like Lorentz, color, or Dirac.
The momentum integrals are defined as
\begin{align}
	\int_q = \int \frac{d^4q}{(2\pi)^4}
\end{align}
and the Fourier transformation from position to momentum space by\footnote{Note that there are some subtleties with respect to the definition of flow of momenta, in particular when fields are complex or Grassmann.
As defined here, momenta in \textit{all} $n$-point functions are defined such that momentum conservation is realized as $\sum_i p_i=0$, e.g., $p+q=0$ in \eref{eq:prop_mom}.}
\begin{align}
	\phi(p) = \int dx \,\phi(x)\,e^{-i\,p\,x}.
\end{align}
The (bare) propagator, as the inverse of the two-point function, is
\begin{align}
D_0(x-y) &= \left(\frac{\de^2 S[\phi]}{\de\phi(x)\de\phi(y)}\right)^{-1}=\left(S_{rs}\right)^{-1}=\left(-\partial^2 + m^2\right)^{-1}\delta(x-y),\\
\label{eq:prop_mom}
D_0(p,q) &= (2\pi)^4\de (p+q)D_0(p), \qquad D_0(p)=\frac{1}{p^2 + m^2}.
\end{align}
Due to translational invariance, the propagator depends only on the difference $x-y$ and momentum is conserved.
As the corresponding delta functional is trivially integrated out in diagrams, we only need $D_0(p)$.

Introducing a source $J$ for the field, the \textbf{generating functional for correlation functions} $\mathbf{Z[J]}$ is defined as
\begin{align}
	Z[J]=\int \mathcal{D}[\phi] e^{-S[\phi] + \phi_j J_j}=:e^{W[J]},
\end{align}
where we also introduced the \textbf{generating functional of connected correlation functions} $\mathbf{W[J]}$.
$\mathcal{D}[\phi]$ is the measure of the path integral.
It can be imagined as integration over all possible field configurations.
Again, the indices in $\phi_j J_j$ are a shorthand for integration (and for summation over potential indices in the case of other fields).
Performing a Legendre transformation of $W[J]$, one obtains the \textbf{effective action} $\mathbf{\Gamma[\Phi]}$, which, or generalizations thereof to be introduced later, is the central object of functional methods.
The Legendre transformation switches the dependence from the sources $J$ to the derivative of $W[J]$ denoted by $\Phi$:
\begin{align}
	\Phi_i\equiv \left\langle \phi_i\right\rangle _{J}=\frac{
	  \delta W}{\delta J_i}=Z[J]^{-1}\int \mathcal{D}[\phi] \,\phi_i\,	 e^{-S + \phi_j J_j} .
\end{align}
The notation $\langle \ldots\rangle_J$ is used to indicate that sources are non-zero here.
Setting them to zero leads to the physical expectation value of the field $\phi$:
$\Phi_{\rm phys}:=\left\langle \phi\right\rangle _{J=0}$.
The effective action is then defined as
\begin{align}\label{eq:Gamma1PI}
	\Gamma[\Phi]&:=\sup_J(-W[J]+J_i \Phi_i ).
\end{align}
Generalizations of the effective action are the $n$PI effective action and the effective average action, see Sec.~\ref{sec:nPI} and Sec.~\ref{sec:FRG}, respectively.

\begin{figure}[tb]
	\includegraphics[width=\textwidth]{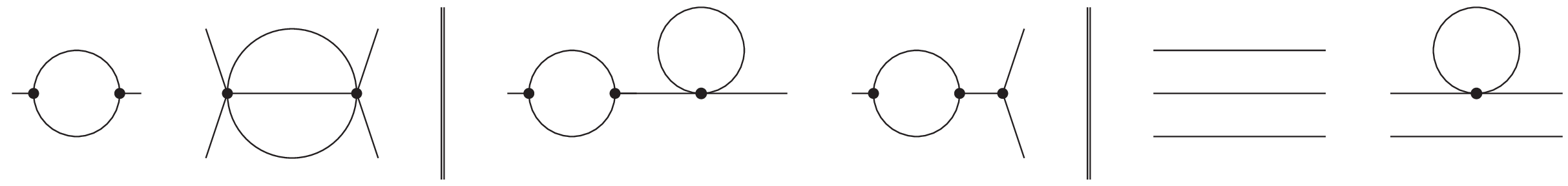}
	\caption{Examples for different diagram types: 1PI (left), connected but one-particle reducible (middle), and disconnected (right).}
	\label{fig:diagramTypes}
\end{figure}

The effective action $\Gamma[\Phi]$ is the generating functional for the one-particle irreducible, or simply 1PI, correlation functions.
One-particle irreducibility refers to the property that they cannot be separated into disconnected pieces by cutting a single line in diagrams.
On the other hand, all other types of correlation functions can be constructed by gluing them together with a propagator (connected but one-particle reducible correlation functions) or simply multiplying them (disconnected correlation functions), see \fref{fig:diagramTypes} for examples.
The 1PI correlation functions, also referred to as $n$-point functions, are given by
\begin{align}\label{eq:Gamma_expansion}
\Gamma_{i_1\ldots i_n}=\Gamma^{J=0}_{i_1\ldots i_n}=(-1)^\kappa\frac{\delta^n \Gamma[\Phi]}{\delta \Phi_{i_1}\ldots\delta \Phi_{i_n}}\Bigg|_{\Phi=\Phi_\text{phys}}.
\end{align}
During the derivation process, when the sources are not set to zero and the fields do not take their expectation values, we deal with field-dependent $n$-point functions $\Gamma^J_{ij\ldots}$:
\begin{align}\label{eq:GammaJ}
 \Gamma^J_{i_1\ldots i_n}=(-1)^\kappa\frac{\delta^n \Gamma[\Phi]}{\delta \Phi_{i_1}\ldots\delta \Phi_{i_n}}.
\end{align}
The sign is determined by $\kappa=0$ for $n=2$ and $\kappa=1$ for $n>2$.
This is merely a conventional choice which entails that the signs of diagrams do not change upon the differentiations discussed below.
 
\begin{figure}[b]
	\includegraphics[width=\textwidth]{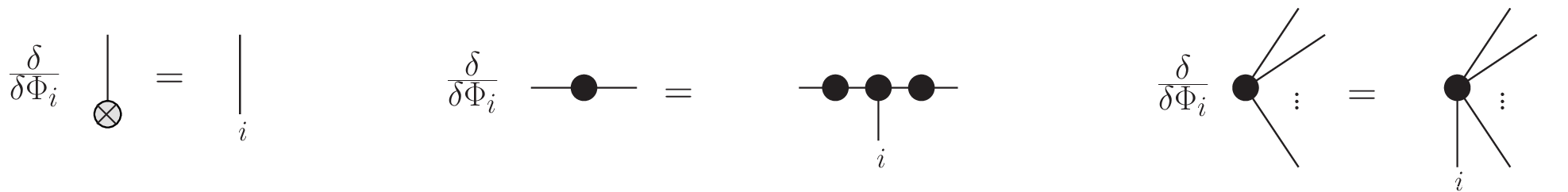}
	\caption{Differentiation rules: derivatives of a field (left), a propagator (middle), and a vertex (right).
	Disks represent (here field-dependent) vertices or propagators, crosses represent fields.}
	\label{fig:diffRules}
\end{figure}

Propagators are the inverse of the two-point functions:
\begin{align}\label{eq:prop}
	D_{ij}:=\frac{\de^2 W[J]}{\de J_i \de J_j}\Bigg|_{J=0}
   =\left[\left(\frac{\delta^{2}\Gamma[\Phi]}{\delta\Phi_i\delta\Phi_j}\right)^{-1}\right]_{ij}\Bigg|_{\Phi=\Phi_\text{phys}}=\left(\Gamma_{ij}\right)^{-1}\Bigg|_{\Phi=\Phi_\text{phys}}.
\end{align}
Equivalently, we call $D^J_{ij}=(\Gamma^J_{ij})^{-1}$ field-dependent propagator.
Its derivative is given by
\begin{align}\label{eq:deriv_D}
\frac{\delta}{\delta\Phi_{i}}D_{jk}^{J}          &=D_{jm}^{J}\Gamma_{imn}^{J}D_{nk}^{J},
\end{align}
which can be obtained from $\de (D^J_{ij}\Gamma^J_{jk})/\de \Phi_l=0$.
Note that the minus sign introduced by convention in the definition of the field dependent $n$-point functions in \eref{eq:GammaJ} canceled a minus sign here, the advantage being that the sign of a diagram does not depend on the number of vertices.
The basic differentiation rules are then $\de (\Phi_j F_{j\ldots})/\de \Phi_i =F_{i\ldots}$, \eref{eq:Gamma_expansion}, and \eref{eq:deriv_D}.
They are shown graphically in \fref{fig:diffRules}.

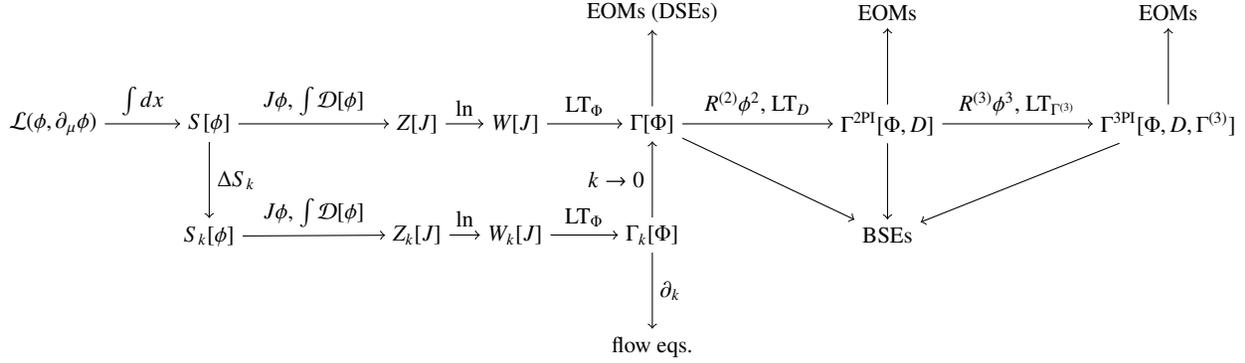
\begin{figure}[tb]
	\begin{tikzpicture}
	\node (L) {$\mathcal{L}(\phi,\partial_\mu \phi)$};
	\node (S) [right=1cm of L] {$S[\phi$]};
	\node (Z) [right=2cm of S] {$Z[J]$};
	\node (W) [right=0.5cm of Z] {$W[J]$};
	\node (Gamma) [right=1cm of W] {$\Gamma[\Phi]$};
	\node (Gamma2PI) [right=2cm of Gamma] {$\Gamma^{\text{2PI}}[\Phi,D]$};
	\node (Gamma3PI) [right=2cm of Gamma2PI] {$\Gamma^{\text{3PI}}[\Phi,D,\Gamma^{(3)}]$};

	\node (Sk) [below=1cm of S] {$S_k[\phi$]};
	\node (Zk) [below=1cm of Z] {$Z_k[J]$};
	\node (Wk) [below=1cm of W] {$W_k[J]$};
	\node (Gammak) [below=1cm of Gamma] {$\Gamma_k[\Phi]$};

	\node (DSEs) [above=1cm of Gamma] {EOMs (DSEs)};
	\node (EOMs2PI) [above=1cm of Gamma2PI] {EOMs};
	\node (EOMs3PI) [above=1cm of Gamma3PI] {EOMs};

	\node (flowEq) [below=1cm of Gammak] {flow eqs.};

	\node (BSE) [below=1cm of Gamma2PI] {BSEs};

	\draw[->] (L) -- node[above] {$\int dx$} (S);
	\draw[->] (S) -- node[above] {$J\phi$, $\int \mathcal{D}[\phi]$} (Z);
	\draw[->] (Z) -- node[above] {$\ln$} (W);
	\draw[->] (W) -- node[above] {LT$_\Phi$} (Gamma);
	\draw[->] (Gamma) -- node[above] {$R^{(2)}\phi^2$, LT$_D$} (Gamma2PI);
	\draw[->] (Gamma2PI) -- node[above] {$R^{(3)}\phi^3$, LT$_{\Gamma^{(3)}}$} (Gamma3PI);
	\draw[->] (Gamma) -- node[above] {} (BSE);
	\draw[->] (Gamma2PI) -- node[above] {} (BSE);
	\draw[->] (Gamma3PI) -- node[above] {} (BSE);

	\draw[->] (S) -- node[right] {$\Delta S_k$} (Sk);
	\draw[->] (Sk) -- node[above] {$J\phi$, $\int \mathcal{D}[\phi]$} (Zk);
	\draw[->] (Zk) -- node[above] {$\ln$} (Wk);
	\draw[->] (Wk) -- node[above] {LT$_\Phi$} (Gammak);

	\draw[->] (Gamma) -- node[above] {} (DSEs);
	\draw[->] (Gamma2PI) -- node[above] {} (EOMs2PI);
	\draw[->] (Gamma3PI) -- node[above] {} (EOMs3PI);

	\draw[->] (Gammak) -- node[right] {$\partial_k$} (flowEq);

	\draw[->] (Gammak) -- node[left] {$k\rightarrow 0$} (Gamma);

	\end{tikzpicture}
	\caption{Relations between different generating functionals and related quantities.
	LT$_A$, $A\in\{\Phi,D,\Gamma^{(3)}\}$ denotes the Legendre transformation with respect to $A$.
	}
	\label{fig:relations}
\end{figure}

\subsection{Equations of motion}

Effective actions contain all information about a theory but in their full form they are complicated objects.
In Sec.~\ref{sec:QFT} we introduced the 1PI effective action.
Below, other $n$PI effective actions will be introduced.
They are all equivalent, but they (can) lead to different equations of motion.
Fig.~\ref{fig:relations} summarizes the relations between the different generating functionals and various functional equations.
We will start with the equations of motion of the 1PI effective action, which are called Dyson-Schwinger equations or Schwinger-Dyson equations \cite{Dyson:1949ha,Schwinger:1951ex,Schwinger:1951hq}.
After introducing $n$PI effective actions with $n>1$, their equations of motion will be discussed.
The derivations of the equations will be sketched for the scalar model with the condensed notation introduced in Sec.~\ref{sec:QFT}.

\subsubsection{Dyson-Schwinger equations}

The path integral formalism is the most convenient and hence also most common way to derive DSEs.
For an alternative using the operator formalism see \cite{Rivers:1988pi}.
We start with the integral over a total derivative which vanishes:
\begin{align}\label{eq:DSE-Z}
  0=&\int D[\phi] \frac{\delta}{\delta \phi_i} e^{-S[\phi] + \phi_j J_j}
      =\int D[\phi] \left( -\frac{\delta S[\phi]}{\delta \phi_i}
      + J_i \right) e^{-S[\phi] + \phi_j J_j}=\left( -\frac{\delta S[\phi']}{
      \delta \phi'_i}\Bigg\vert_{\phi'_i=\delta/\delta J_i} +J_i \right) Z[J]\,.
\end{align}
In the last step, the derivative was pulled in front of the integral which is possible by changing the argument from the field $\phi$ to the derivative with respect to the source $J$.
Using the relation
\begin{align}
	e^{-W[J]}\left(\frac{\delta}{\delta J_i}\right)e^{W[J]}=
	\frac{\delta W[J]}{\delta J_i}+\frac{\delta}{\delta J_i},
  \end{align}
we can switch in \eref{eq:DSE-Z} to the generating functional of connected correlation functions, $W[J]$:
\begin{align}
  -\frac{\delta S[\phi']}{\delta \phi'_i}\Bigg\vert_{\phi'_i=
  \frac{\delta W[J]}{\delta J_i}
  +\frac{\delta}{\delta J_i}} +J_i=0.
\end{align}
Finally, by performing the Legendre transformation from \eref{eq:Gamma1PI}, we obtain the master equation for 1PI correlation functions:
\begin{align}\label{eq:DSE-master}
  \frac{\delta \Gamma[\Phi]}{\delta \Phi_i}=\frac{\delta S[\phi']}{\delta
  \phi'_i}\Bigg\vert_{\phi'_i=\Phi_i+D_{ij}^J  \, \frac{\delta}{\delta \Phi_j}}.
\end{align}
Equations of motion for $n$-point functions are obtained by applying $n-1$ further derivatives and setting the sources to zero.
More details can be found in \cite{Alkofer:2008nt,Swanson:2010pw,Huber:2011qr,Huber:2018ned,Huber:2019dkb}.

\begin{figure}[b]
	\centering
	\includegraphics[width=0.98\textwidth]{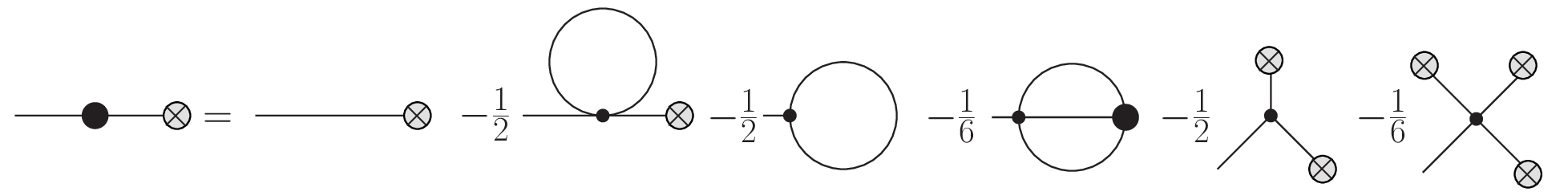}
		\caption{Graphical representation of \eref{eq:dse_1deriv}.
		Internal propagators are fully dressed.
		Crosses correspond to fields $\Phi$.
		Dots represent bare and disks dressed vertices.}
		\label{fig:dse_scalar_1deriv}
\end{figure}

\begin{figure}[b]
	\centering
		\includegraphics[width=0.98\textwidth]{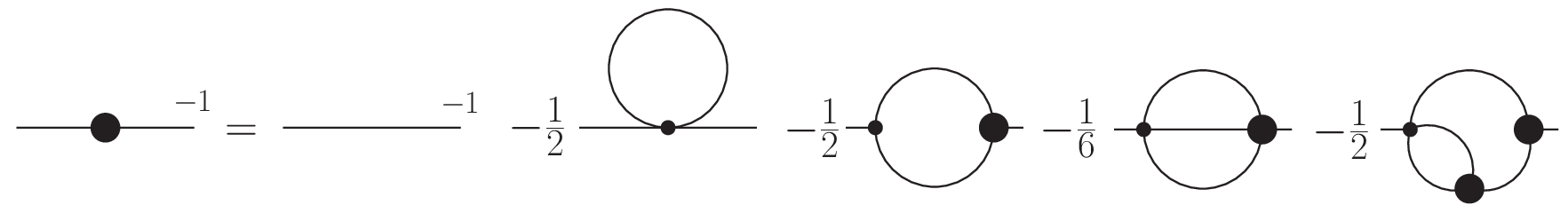}
		\caption{The DSE for the propagator of the scalar theory from \eref{eq:scalarLagrangian}.}
		\label{fig:dse_scalar_prop}
\end{figure}

Let us consider the scalar theory in \eref{eq:scalarLagrangian} as an example.
The derivative of the effective action, as written in \eref{eq:scalarAction2}, is
\begin{align}
\frac{\delta S[\phi]}{\delta \phi_i} = S_{ij}\phi_j - \frac{1}{2}S_{ijk}\phi_j \phi_k - \frac{1}{3!}S_{ijkl}\phi_j \phi_k \phi_l.
\end{align}
Replacing the fields as $\phi \rightarrow \Phi + D^J \de/\de \Phi$ yields
\begin{align}
\frac{\delta \Gamma[\Phi]}{\delta \Phi_i}=\frac{\delta S[\phi']}{\delta \phi'_i}\Bigg\vert_{\phi'_i=\Phi_i+D_{ij}^J  \, \frac{\delta}{\delta \Phi_j}} &= S_{ij}\Phi_j - \frac{1}{2}S_{ijk}\left(\Phi_j + D_{jl}^J \frac{\delta}{\delta \Phi_l}\right)\Phi_k - \frac{1}{3!}S_{ijkl}\left(\Phi_j + D_{jm}^J \frac{\delta}{\delta \Phi_m}\right)\left(\Phi_k + D_{kn}^J \frac{\delta}{\delta \Phi_n}\right)\Phi_l\nonumber\\
&= S_{ij}\Phi_j - \frac{1}{2}S_{ijk}\left(\Phi_j \Phi_k + D_{jk}^J\right) - \frac{1}{3!}S_{ijkl}\left(\Phi_j \Phi_k \Phi_l + 3D_{jk}^J \Phi_l + D_{jm}^J D_{kn}^J D_{lo}^J \Gamma^J_{mno} \right).\label{eq:dse_1deriv}
\end{align}
The symmetry of $S_{ijkl}$ under exchange of indices was used.
This is graphically represented in \fref{fig:dse_scalar_1deriv}.
Performing a derivative with respect to $\Phi_j$ and setting the sources to zero, the left-hand side becomes the dressed two-point function (or equivalently the inverse propagator).
The right-hand side contains the bare two-point function and one- and two-loop diagrams.
\begin{align}
\left(D_{ij}\right)^{-1}=\frac{\delta \Gamma[\Phi]}{\delta \Phi_i\delta \Phi_j}\Bigg|_{\Phi=\phi_\text{phys}} &= S_{ij} - \frac{1}{2}S_{ikl}D_{lm}D_{kn}\Gamma_{mnj} - \frac{1}{2}S_{iklj}D_{kl} - \frac{1}{2!}S_{iklm}D_{kn}D_{lo}D_{mp}D_{rs}\Gamma_{opr}\Gamma_{nsj} -\frac{1}{3!}S_{iklm} D_{kn}D_{lo}D_{mp}\Gamma_{nopj} .
\end{align}
This equation is shown in \fref{fig:dse_scalar_prop}.

\subsubsection{$n$PI effective actions and their equations of motion}
\label{sec:nPI}

The 1PI effective action introduced in Sec.~\ref{sec:QFT} depends only on the field $\Phi$ and is obtained by a single Legendre transformation.
By introducing sources for the propagators and vertices one can perform further Legendre transformations to obtain $n$PI effective actions.
The 2PI effective action \cite{Cornwall:1974vz} has then a functional dependence not only on the field $\Phi$ but also on the full propagator $D$.
$n$PI effective actions with $n>2$ depend in addition on the vertices of the theory with up to $n$ legs \cite{Berges:2004pu,Carrington:2010qq}.
Without approximations, all $n$PI effective actions, $n>0$, are equivalent, but in practical calculations, their equations of motion may differ under approximations.
In this context, $n$-particle irreducibility means that one cannot separate the diagrams into disconnected loop diagrams by cutting at most $n$ lines.
However, this only holds for $n\leq4$ \cite{Carrington:2010qq} and the $n$ should thus be interpreted as highest dynamically included $n$-point function.
As we will see, $n$PI effective actions lead to similar equations for propagators and vertices as Dyson-Schwinger equations but there are some decisive differences, in particular with regard to the systematics of truncations.

To discuss the most relevant features of $n$PI effective actions, let us consider the 3PI effective action as an example.
By introducing sources $R^{(2)}_{ij}$ and $R^{(3)}_{ijk}$ for quadratic and cubic terms, the path integral becomes
\begin{align}
 e^{W[J,R^{(2)},R^{(3)}]} = Z[J,R^{(2)},R^{(3)}]=\int D[\phi] e^{-S + \phi_i J_i + \frac{1}{2} R^{(2)}_{ij}\phi_i \phi_j + \frac{1}{3!} R^{(3)}_{ijk}\phi_i \phi_j \phi_k }.
\end{align}
With $W = W[J, R^{(2)}, R^{(3)}]$ it follows
\begin{align}
 \frac{\de W}{\de J_i}=\langle \phi_i \rangle_J = \Phi_i,\quad
 \frac{\de W}{\de R^{(2)}_{ij}}=\frac{1}{2}\left( D_{ij}^J  + \Phi_i \Phi_j \right),\quad
 \frac{\de W}{\de R^{(3)}_{ijk}}=\frac{1}{3!}\left( D^{(3),J}_{ijk} + D_{ij}^J \Phi_k + D_{ik}^J \Phi_j + D_{jk}^J \Phi_i  + \Phi_i \Phi_j \Phi_k \right),
\end{align}
where $D^{(3),J}_{ijk}$ is the third derivative of $W$ with respect to $J$.
The 3PI effective action is obtained by a triple Legendre transformation as
\begin{align}\label{eq:nPI-action}
 &\Gamma = \Gamma[\Phi,D,\Gamma^{(3)}]=  - W + \frac{\de W}{\de J_i} J_i + \frac{\de W}{\de R^{(2)}_{ij}} R^{(2)}_{ij} + \frac{\de W}{\de R^{(3)}_{ijk}} R^{(3)}_{ijk}.
\end{align}
In turn,
\begin{align}\label{eq:3PI-derivs}
 \frac{\de \Gamma}{\de \Phi_i} &= J_i, \qquad \frac{\de \Gamma}{\de D^{(2)}_{ij}} = \frac{1}{2} R^{(2)}_{ij}, \qquad \frac{\de \Gamma}{\de \Gamma^{(3)}_{ijk}} = \frac{1}{6} D_{ai}D_{bj}D_{kc}R^{(3)}_{abc}.
\end{align}
For the last equation, $D^{(3),J}_{ijk}=D_{ai}^JD_{bj}^JD_{ck}^J\Gamma^{(3),J}_{ijk}$ was used, see \eref{eq:deriv_D}.
The equations of motion for the propagators and three-point functions are a direct consequence of the stationarity conditions obtained by setting the sources to zero in \eref{eq:3PI-derivs}:
\begin{align}\label{eq:stationarity_conds}
 \frac{\delta \Gamma}{\de \Phi_i}\Bigg|_{J=R^{(2)}=R^{(3)}=0}=0, \qquad \frac{\delta \Gamma}{\de D_{ij}}\Bigg|_{J=R^{(2)}=R^{(3)}=0}=0, \qquad \frac{\delta \Gamma}{\de \Gamma^{(3)}_{ijk}}\Bigg|_{J=R^{(2)}=R^{(3)}=0}=0.
\end{align}

\begin{figure}[tb]
  \includegraphics[height=1.7cm]{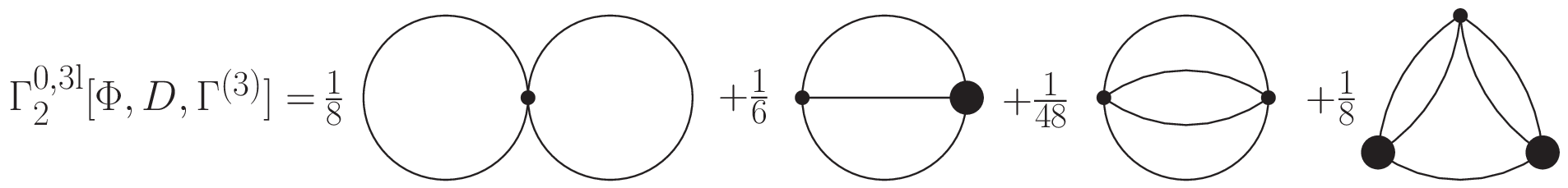}\\
  \vskip3mm
  \includegraphics[height=1.7cm]{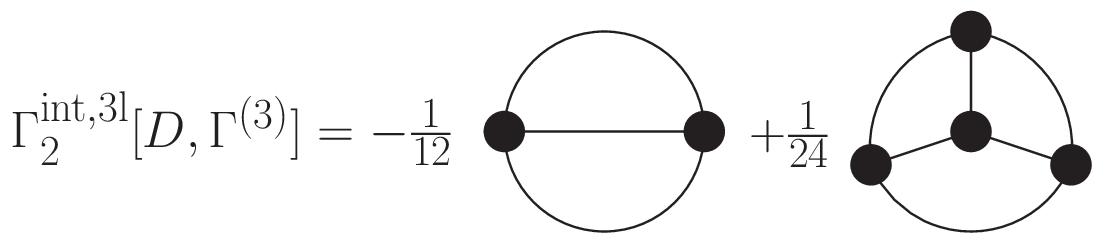}
 \caption{$\Gamma_2^{0,\text{3l}}[\Phi,D,\Gamma^{(3)}]$ and $\Gamma_2^{\text{int},\text{3l}}[D, \Gamma^{(3)}]$ of the 3PI effective action at three-loop order.}
 \label{fig:Gamma2}
\end{figure}

To exemplify the equations of motion, we consider the three-loop expansion for a theory with cubic and quartic interactions, see \eref{eq:scalarLagrangian}.
The loop expansion will be discussed further in Sec.~\ref{sec:loopExp}.
Since we will only discuss the structure of the equations, this can directly be generalized to other cases and even fermionic fields if the signs from fermionic loops and the modified symmetry factors are taken into account.
Starting with the corresponding loop expansion of the 1PI effective action $\Gamma[\Phi]$, the 3PI effective action is obtained via two more Legendre transformations as \cite{Berges:2004pu}
\begin{subequations}\label{eq:3l-3PI}
	\begin{align}
 \Gamma^\text{3l}[\Phi, D, \Gamma^{(3)}] &= S[\Phi] + \frac{1}{2}\ln D^{-1}_{ii} + \frac{1}{2} S_{ij}[\Phi] D_{ji} - \Gamma_2^{\text{3l}}[\Phi,D, \Gamma^{(3)}],\\
 \Gamma_2^{\text{3l}}[\Phi,D, \Gamma^{(3)}] &= \Gamma_2^{0,\text{3l}}[\Phi,D, \Gamma^{(3)}] + \Gamma_2^{\text{int},\text{3l}}[D, \Gamma^{(3)}],
	\end{align}
\end{subequations}
where $S_{ij}[\Phi]$ is the field dependent, viz. the sources are not set to zero, inverse propagator defined as $\de^2 S[\Phi]/\de\Phi_i\de\Phi_j$.
The last term is split into a part depending on bare vertices and a part depending on dressed quantities only.
Both parts are depicted graphically in \fref{fig:Gamma2}.

The equation of motion for the propagator follows from the stationarity condition $\de \Gamma/\de D|_0=0$ which leads to
\begin{align}
 D^{-1}=D^{-1}_0 - 2 \frac{\delta \Gamma_2^{\text{3l}}}{\de D}\Bigg|_0,
\end{align}
where $(D_0)_{ij}=(S_{ij})^{-1}_{J=0}$ is the bare propagator and the $|_0$ indicates that all sources are zero and thus fields, propagators and vertices are replaced by their solutions.
The second term on the right-hand side is the selfenergy, $\Sigma = -2\de\Gamma_2^{\text{3l}}/\de D|_0$.
The derivative in the last term corresponds to cutting propagator lines in the diagrams of $\Gamma_2^{\text{3l}}$ and leads to
\begin{align}\label{eq:3PI-prop_not_resummed}
	\includegraphics[width=0.8\textwidth]{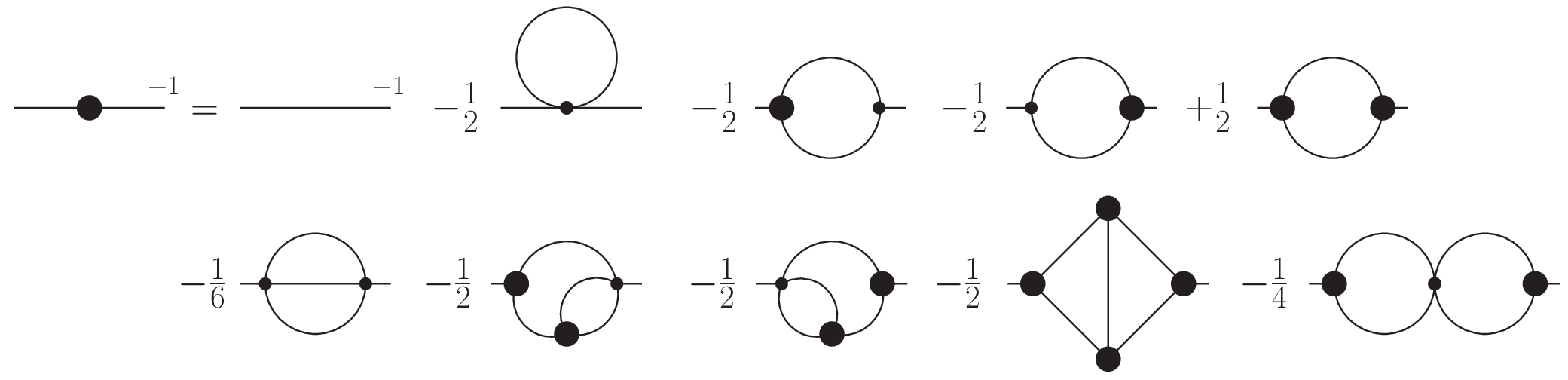}.
\end{align}
Note that the derivative was symmetrized to obtain the manifestly symmetric form of the equation. 
This equation has some similarities to the corresponding DSE, see \fref{fig:dse_scalar_prop}, but there are additional diagrams and the prefactors are different.
Using the equation of motion for the three-point function, the connection to the DSE becomes more obvious, though.
This equation is derived from the stationarity condition $\de\Gamma_2^{\text{3l}}/\de \Gamma^{(3)}|_0=0$.
The derivative corresponds to cutting out three-point functions in $\Gamma_2^{\text{3l}}$ which yields
\begin{align}\label{eq:3PI-vert}
  \includegraphics[height=1.9cm]{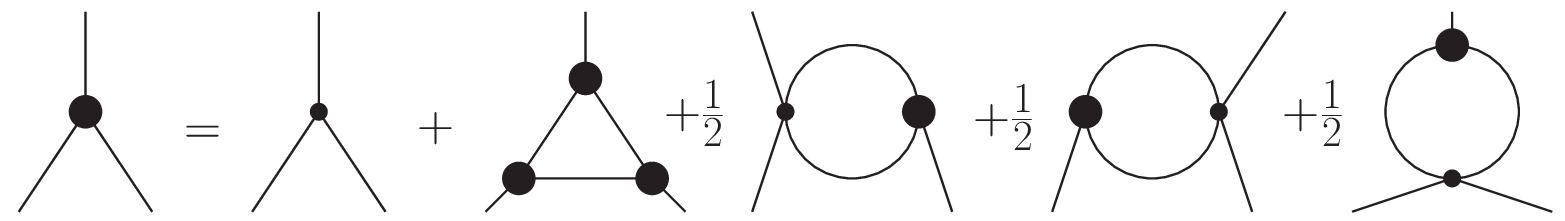}.
\end{align}
From this equation it becomes obvious why at least a three-loop expansion is needed for a consistent inclusion of the three-point function.
If we had used a two-loop expansion, the three-point function would just be bare which is equivalent to a two-loop expansion of the 2PI effective action.
In general, there is an equivalence hierarchy of loop expansions between different $n$PI effective actions \cite{Berges:2004pu} and we need at least an $n$-loop expansion for an $n$PI effective action.
Using \eref{eq:3PI-vert} for the right three-point function in the fourth one-loop diagram in \eref{eq:3PI-prop_not_resummed}, the two-point equation can be rewritten as
\begin{align}\label{eq:3PI-prop}
	\includegraphics[width=0.9\textwidth]{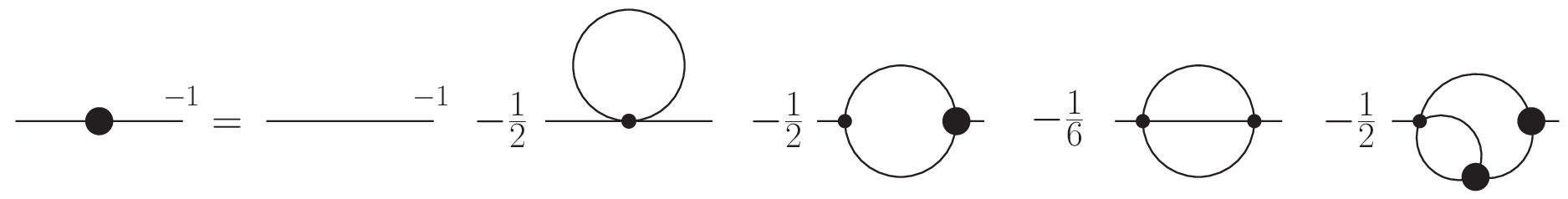}.
\end{align}
This form resembles the corresponding DSE except for one bare four-point function.
Using a 4PI effective action, the two-point equation would actually completely agree with the DSE and thus look like an exact equation despite the loop expansion \cite{Carrington:2010qq}.
It should be noted that within the truncation the vertices are not exact, though, so formally the propagator equation is not exact either, despite its suggestive form.

$n$PI effective actions can also be used to derive kernels for bound state equations.
This will be described in Sec.~\ref{sec:BSEs}.

\subsection{Bound state equations}
\label{sec:BSEs}

Bound states are key objects studied in quantum field theories.
In particular in the case of QCD, they are primary observables as the elementary quarks and gluons are not directly observable due to confinement.\footnote{Note that conceptually also for the rest of the standard model we are dealing with bound states described by gauge invariant states. The Fr\"ohlich-Morchio-Strocchi mechanism explains the success of conventional methods employing elementary fields to calculate masses \cite{Frohlich:1980gj,Frohlich:1981yi,Maas:2023emb}.}
Consequently, a lot of effort has been invested in developing theoretical methods to calculate them.
Most notably, lattice QCD constitutes a successful and highly developed first-principle approach to this problem, see, e.g., \cite{Prelovsek:2025gmd}.
Functional methods also provide a powerful framework for this purpose, in particular relying on bound state equations.
This section will provide a short introduction to the main features of bound state equations.
More details can be found in \cite{Eichmann:2025wgs}.

\begin{figure}[tb]
 \includegraphics[height=2cm]{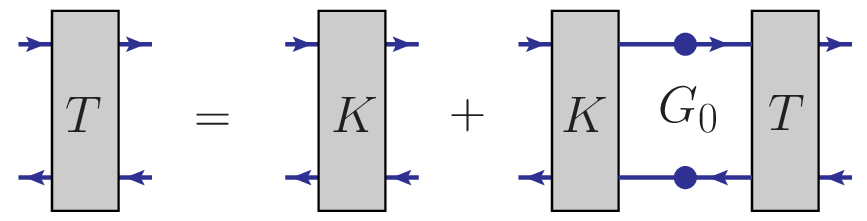}\\
 \vskip2mm
 \includegraphics[height=2cm]{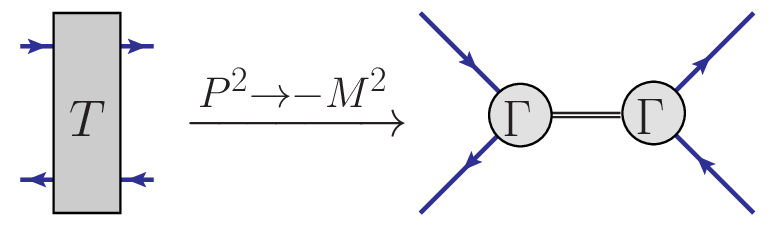}\hfill
 \includegraphics[height=2cm]{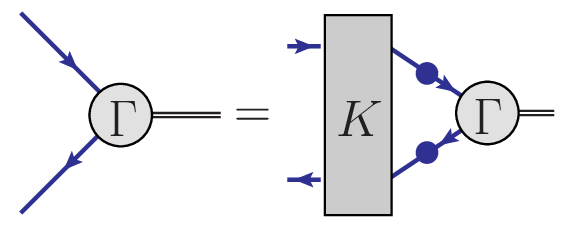}
 \caption{Top: Dyson equation for the scattering matrix $T$.
 Bottom Left: The scattering matrix $T$ at the pole $P^2=-M^2$.
 Bottom right: Homogeneous BSE for a meson.}
 \label{fig:TDysonBSE}
\end{figure}

The spectrum of a quantum field theory is encoded in the so-called scattering matrix $T$ which describes the nontrivial part of the $S$-matrix operator connecting the incoming to the outgoing states of a scattering process, see, e.g., \cite{Mai:2025wjb}.
It can be formally written in form of a Dyson equation,
\begin{align}\label{eq:T_Dyson}
T=K+K\,G_0\,K+K\,G_0\,K\,G_0\,K+\ldots=K(\mathbb{1}+G_0\,T).
\end{align}
$G_0$ is the propagator of all particles and $K$ the $n$-particle irreducible scattering kernel, viz., it contains all possible interactions except those obtained by iterating the equation.
To be concrete, let us consider the case of a two-body bound state, for example, a quark and an antiquark forming a meson.
The corresponding Dyson equation is shown in \fref{fig:TDysonBSE} where $G_0$ is the propagator of a quark and an antiquark, $K$ the 2-particle irreducible interaction between them, and the scattering matrix $T$ the amputated, connected part of the four-quark function.
A meson corresponds to a pole in the scattering matrix which takes the following form in its vicinity, see also \fref{fig:TDysonBSE}:
\begin{align}
  T\xrightarrow{P^2\rightarrow -M^2} \frac{\Gamma\,\overline{\Gamma}}{P^2+M^2}.
\end{align}
$P$ is the total momentum of the meson of mass $M$.
$\Gamma$ is called the Bethe-Salpeter amplitude and $\overline{\Gamma}$ is its charge conjugate.
Let us recall that in the Euclidean metric we use here, the pole condition is $P^2=-M^2$.
Close to $P^2=-M^2$ the pole contribution dominates in the Dyson equation (\ref{eq:T_Dyson}) and one can extract the homogeneous Bethe-Salpeter equation \cite{Salpeter:1951sz,Salpeter:2008sp}
\begin{align}\label{eq:BSE}
 \Gamma(p,P)=\int_q \, K(p,q,P)\,G_0(q,P)\,\Gamma(q,P)
\end{align}
where the momentum dependence is formally indicated and all indices are implicit.
Fig.~\ref{fig:TDysonBSE} shows this equation graphically.
$p$ and $q$ are relative momenta of the amplitudes on the left- and right-hand side.
This is a homogeneous Fredholm integral equation of second kind.
The BSE (\ref{eq:BSE}) can be rewritten with the Bethe-Salpeter wave function $\Psi=G_0\,\Gamma$ as
\begin{align}\label{eq:BSE-wf}
 G_0(p,P)\,\Gamma(p,P)=\Psi(p,P)=\int_q \, G_0(p,P)\, K(p,q,P)\,\Psi(q,P).
\end{align}
The discussion above can be generalized to three and more constituents.
In particular, three-body bound state equations, called Faddeev equations, describing baryons have been widely investigated \cite{Eichmann:2016yit} and also for four-quark states many results exist \cite{Eichmann:2020oqt}.
However, the more constituents there are, the more complicated the equations become, see \cite{Eichmann:2025gyz} for the current state of the art with regard to five constituents.

Up to now we did not specify the quantum numbers of a specific meson.
These are encoded in the Bethe-Salpeter amplitude $\Gamma$.
For total angular momentum $J$, which is an integer number for mesons, it can be expanded in a tensor basis as
\begin{align}
 [\Gamma(p,P)]_{\alpha\beta}^{\mu_1\ldots\mu_J} = \sum_i f_i(p^2,p\cdot P,P^2)\,[\tau_i(p,P)]_{\alpha\beta}^{\mu_1\ldots\mu_J},
\end{align}
where $\mu_i$ are $J$ Lorentz indices and $\alpha$ and $\beta$ are Dirac indices.
The tensors $\tau_i$ are constructed in accordance with the desired quantum numbers of total angular momentum $J$ and parity $\mathsf{P}$.
By a proper choice of the basis $\tau_i$, charge parity $\mathsf{C}$ can be shifted completely into the symmetry properties of the amplitude under charge conjugation, see \cite{Eichmann:2016yit} for details.

As a simple example, the basis for a pseudoscalar meson can be spanned by the following four tensors:
\begin{align}
 \tau_1(p,P) = \gamma_5,\quad \tau_2(p,P) = i\,\slashed{P}\gamma_5,\quad \tau_3(p,P) = i\,p\cdot P\,\slashed{p}\,\gamma_5, \quad \tau_4(p,P) = [\slashed{p},\slashed{P}]\,\gamma_5.
\end{align}
The scalar product $p\cdot P$ was introduced to make the basis element $\tau_3$ invariant under charge conjugation.
The form factors $f_i$ even in $p\cdot P$ correspond then to pseudoscalar mesons with $J^{\mathsf{PC}}=0^{-+}$, while those odd in $p\cdot P$ correspond to exotic mesons with $J^{\mathsf{PC}}=0^{--}$.
In practical calculations, the selection of states with positive or negative charge parity can be done by a Chebyshev expansion in the angle between $p$ and $P$ and by selecting only even or odd coefficients, respectively, see, for instance, \cite{Sanchis-Alepuz:2017jjd}.

The central element for a BSE is the scattering kernel $K$.
$n$PI effective actions provide a systematic way for its derivation \cite{Fukuda:1987su,Komachiya:1989kc,McKay:1989rk}, see also \cite{Sanchis-Alepuz:2015tha,Huber:2020ngt} for applications to modern hadron physics.
In the following, we consider a two-body bound state for the scalar theory discussed in Sec.~\ref{sec:nPI} to illustrate the procedure.
A solution $D^{(0)}$ of the stationarity condition $\delta \Gamma/\delta D|_0=0$ for the propagator is stable if $D^{(1)}=D^{(0)}+\Delta D$, where $\Delta D$ is an infinitesimal perturbation, is also a solution.
From the expansion of the stationarity condition around $D^{(0)}$, this leads to the condition
\begin{align}\label{eq:stability}
 \frac{\delta^2  \Gamma^{\text{3l}}[\Phi, D, \Gamma^{(3)}]}{\delta D_{ij}\delta D_{kl}}\Bigg|_{0} \Delta D_{kl}=0.
\end{align}
The notation is the same as introduced in Sec.~\ref{sec:QFT} where repeated indices represent integration and summation if applicable.
To illustrate how this leads to a BSE, let us consider the three-loop truncated 3PI effective action in \eref{eq:3l-3PI}.
The first derivative leads to
\begin{align}
 \frac{\delta  \Gamma^{\text{3l}}[\Phi, D, \Gamma^{(3)}]}{\delta D_{ij}}=-\frac{1}{2}D^{-1}_{ij}+\frac{1}{2}S_{ij} - \frac{\delta \Gamma_2^{\text{3l}} [\Phi, D, \Gamma^{(3)}]}{\de D_{ij}}.
\end{align}
At this point it is important that we continue with this expression and do not use the equation of motion for the vertex which led to the simplification of the propagator equation in \eref{eq:3PI-prop}.
The second derivative yields\footnote{Note that $\frac{\de D_{kl}^{-1}}{\de D_{ij}}=-D^{-1}_{ki}D^{-1}_{jl}$ follows from $\de (D_{ke}^{-1}D_{el})/\de D_{ij}=0$.}
\begin{align}
 \frac{\delta^2  \Gamma^{\text{3l}}[\Phi, D, \Gamma^{(3)}]}{\delta D_{ij} \delta D_{kl}}=
 \frac{1}{2}D^{-1}_{ki}D^{-1}_{jl} -  \frac{\delta^2 \Gamma_2^{\text{3l}}}{\de D_{ij} \de D_{kl}}.
\end{align}
Plugging this into \eref{eq:stability}, we obtain
\begin{align}
 \left(D^{-1}_{ki}D^{-1}_{jl} - 2 \frac{\delta^2 \Gamma_2^{\text{3l}}}{\de D_{ij} \de D_{kl}}\right)\Bigg|_{0}  \Delta D_{kl}=0.
\end{align}
Identifying  $\Delta D$ with the Bethe-Salpeter wave function $\Psi=G_0\,\Gamma$, we can rewrite this to
\begin{align}
 K_{ij;kl}\,  \Psi_{kl} = \Gamma_{ij}
\end{align}
where the kernel $K$ is defined as
\begin{align}\label{eq:kernel}
	K_{ij;kl}=2\frac{\de^2 \Gamma^{\text{3l}}_2}{\de D_{ij}\de D_{kl}}\Bigg|_{0}.
\end{align}
Let us work this out in more detail.
Since derivatives with respect to the propagator correspond to cutting propagator lines, the kernel is obtained from \eref{eq:3PI-prop_not_resummed} as
\begin{align}
\includegraphics[width=0.9\textwidth]{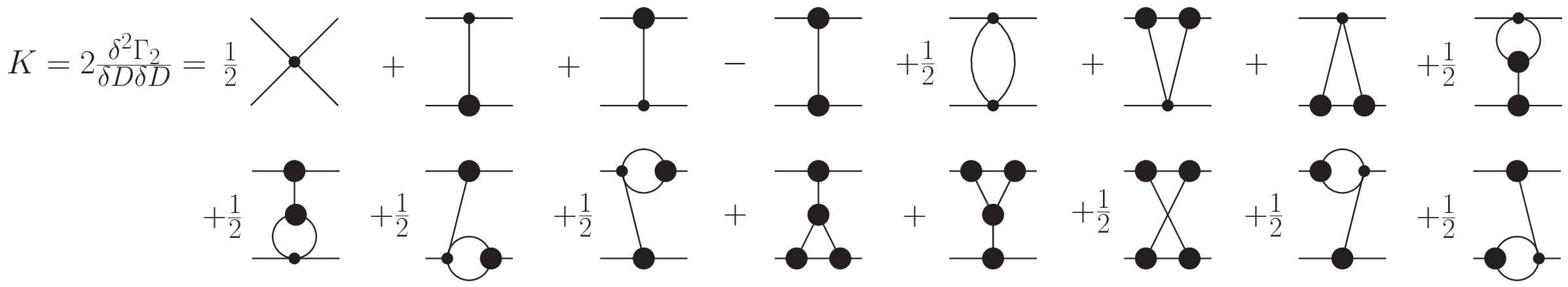}.
\end{align}
We use the equation of motion for the vertex, \eref{eq:3PI-vert}, to replace the bare three-point functions in the second and third diagrams on the right-hand side.
This leads to some cancellations and the final kernel is
\begin{align}\label{eq:3PI-kernel}
\includegraphics[width=0.75\textwidth]{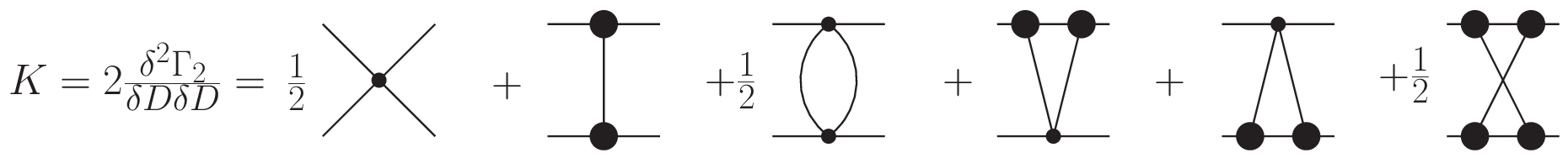}.
\end{align}
Equations (\ref{eq:3PI-vert}), (\ref{eq:3PI-prop}) and (\ref{eq:3PI-kernel}) correspond to a consistent truncation of the correlation functions and the bound state equation.
Most importantly, the truncation can be improved systematically by including more loops in the 3PI effective action.
The kernel $K$ differs from widely used one-particle exchange kernels in particular because of the appearance of loop terms which make its use technically more demanding.
For QCD, corresponding kernels for mesons and glueballs can be obtained directly from \eref{eq:3PI-kernel} by replacing the scalars by quarks, gluons and ghosts.
The Grassmann nature of quarks and ghosts is taken into account by adding minus signs for closed loops and adapting the symmetry factors.

To conclude this section on bound state equations, let us see where the so-called rainbow-ladder truncation, which is widely used in QCD for mesons, see \cite{Eichmann:2025wgs}, fits into this.
It is closely related to a two-loop truncated 2PI effective action, which can easily be obtained from the expression here by reducing the number of loops by one.
This makes the three-point vertex bare.
The four-point interaction is removed because there is no corresponding bare vertex in QCD.
Then, the propagator contains only one diagram (the second one-loop diagram in \eref{eq:3PI-prop}) as does the scattering kernel (the one-particle exchange diagram).
Finally, to make this truncation phenomenologically viable, an effective interaction is introduced.
The 3PI setup can thus be seen as a systematic extension of the rainbow-ladder truncation.

\subsection{Functional renormalization group}
\label{sec:FRG}

For the functional renormalization group, the concept of the so-called effective average action or scale-dependent effective action $\Gamma_k[\Phi]$ is introduced by using a momentum dependent regulator.
It depends on an artificial momentum scale $k$ which allows interpolating between the UV and the IR.
The original idea is that in the limit $k\rightarrow k_\text{in}$, where $k_\text{in}$ is an initial scale, the effective average action corresponds to the classical action, $\Gamma_{k_\text{in}}[\Phi] \rightarrow S[\Phi]$ \cite{Wetterich:1992yh,Dupuis:2020fhh}.
Typically, $k_\text{in}$ can be chosen as the UV cutoff of the theory.
Lowering the scale $k$, quantum fluctuations above $k$ are integrated out successively until for $k\rightarrow 0$ the full quantum effective action $\Gamma[\Phi]$ is recovered.

\begin{figure}[b]
\includegraphics[width=0.49\textwidth]{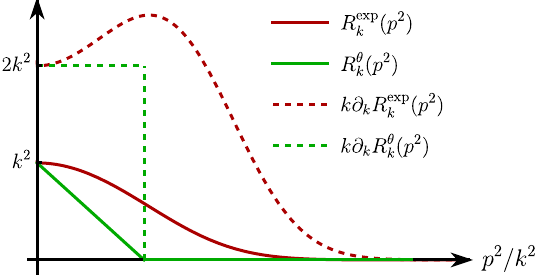}\hfill
\includegraphics[width=0.35\textwidth]{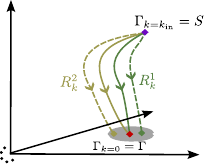}
\caption{Left: Regulator shape functions and their $k$-derivatives for a theta regulator and a variant of an exponential regulator ($m=2$).
Right: Renormalization group trajectories in theory space for two different regulators.
For a given action $S$ the end points would be the same if no truncations were made (continuous lines) although the flow trajectories differ.
Due to truncations (dashed lines), the end points depend on the regulator.
The axes correspond to different operators spanning the theory space which is infinite dimensional as indicated by the dots representing all other operators.}
\label{fig:FRG_regulator_theorySpace}
\end{figure}

This integrating out of momentum modes is realized by a regulator term added to the action $S[\phi]$:
\begin{align}
W_k[J]=\ln \,Z_k[J]=\ln \int D\phi \, e^{-S[\phi] +\int J\, \phi-\frac{1}{2}\int\phi\, R_{k}\, \phi}
\end{align}
For the integral over position or momentum space a generic notation is used.
The regulator function $R_{k}$, which depends on $k$ and a momentum to be denoted $p$ below, needs to respect the following properties:
\begin{enumerate}
\item It has to vanish for $k \rightarrow 0$ to obtain the full effective action in the IR, $\Gamma_{k=0}[\Phi]=\Gamma[\Phi]$.
\item  It has to diverge for $k =k_\text{in}$ at which the microscopic action is defined, $\Gamma_{k=k_\text{in}}[\Phi]=S[\Phi]$; see \cite{Dupuis:2020fhh} for a more detailed discussion.
\item  For small momenta $p^2<k^2$ it must be proportional to $k^2$, thus behaving like an effective mass acting as an IR cutoff for fluctuations with small momenta.
\item  It has to vanish for large momenta $p^2>k^2$ so that it does not interfere with the large momentum behavior.
\end{enumerate}
The qualitative momentum dependence of the regulator function is shown in \fref{fig:FRG_regulator_theorySpace}.
Common examples for regulator functions are the theta (or Litim) regulator \cite{Litim:2001up} and the exponential regulator \cite{Wetterich:1992yh}, respectively:
\begin{align}\label{eq:Gammak_Legendre}
R_k^\theta(p^2)=(k^2-p^2)\theta(k^2-p^2), \qquad R_k^\text{exp}(p^2)=p^2\frac{\left(\frac{p^2}{k^2}\right)^{m-1}}{e^{\left(\frac{p^2}{k^2}\right)^{m}}-1}.
\end{align}
$m$ is a parameter that can be varied to change the shape of the latter.
The former does not only have favorable properties with regard to optimization of the flow equations \cite{Litim:2001up,Pawlowski:2005xe} but also enables analytic calculations in some cases.

\begin{figure}[t]
 \centering
	\includegraphics[width=0.25\textwidth]{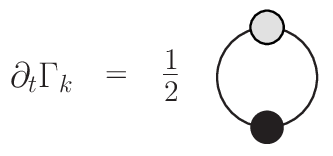}
	\hskip15mm
	\includegraphics[height=2cm]{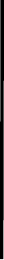}
	\hskip15mm
	\includegraphics[width=0.55\textwidth]{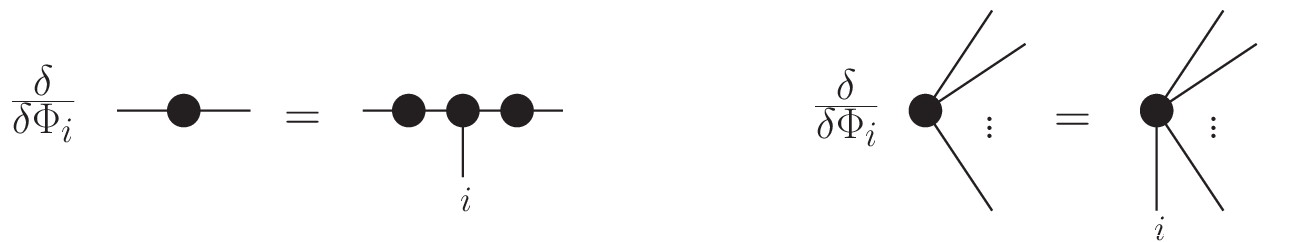}
 \caption{
 Left: The flow equation (\ref{eq:flow_eq}).
 The gray disk denotes the regulator insertion $\partial_t R_k$.
 The dressed propagator $\left(\Gamma_k^{(2)}[\Phi] + R_{k}\right)^{-1}$ is explicitly depicted by a black disk.
 Right: Differentiation rules for a propagator and a vertex, respectively, for the FRG.
 Disks represent (here field-dependent) vertices or propagators.
 }
 \label{fig:fRG_master_prop_diffRules}
\end{figure}

The effective average action is obtained via a modified Legendre transformation of $W_k[J]$:
\begin{align}\label{eq:Gamma_k_Phi-k}
 \Gamma_k[\Phi]=\sup_J \left( -W_k[J]+\int  J \,\Phi - \frac1{2}\int \Phi \,R_{k}\, \Phi \right), \quad \quad \Phi=\frac{\delta W_k[J]}{\delta J}=\langle \phi \rangle_J,
\end{align}
Its dependence on the scale $k$ is described by the so-called Wetterich equation \cite{Wetterich:1992yh}:
\begin{align}\label{eq:flow_eq}
 \partial_t \Gamma_k[\Phi]=& \frac1{2} \text{Tr}\left\{\left(\Gamma_k^{(2)}[\Phi] + R_{k}\right)^{-1} \partial_t R_{k} \right\}.
\end{align}
As often done, the (negative) RG "time" $t=\ln(k/\Lambda)$ is used instead of $k$ with $\partial_t= k\,\partial_k=k\,\frac{\partial}{\partial k}$.
The derivative is sometimes denoted by $\dot{\Gamma}_k[\Phi]$.
Furthermore, a condensed notation is used where $\Gamma_k^{(2)}[\Phi]$ is the second, field-dependent derivative of the effective average action and the trace corresponds to integration in position/momentum space and summation over internal indices as well as over all fields (if $\Phi$ represents more than one field).
This equation is shown diagrammatically in \fref{fig:fRG_master_prop_diffRules}.
For its derivation see the box below.

Eq.~(\ref{eq:flow_eq}) is a one-loop equation with the propagator given by $\left(\Gamma_k^{(2)}[\Phi] + R_{k}\right)^{-1}$.
Solving this flow equation corresponds to integrating out all fluctuations and going from a microscopic description, determined by the classical action $S$, to a macroscopic description.
It should be noted that the trajectories in theory space, the space of all invariant operators of the fields, but not the endpoint depend on the regulator function $R_k$ in the exact equation~(\ref{eq:flow_eq}).
However, this equation can in general not be solved exactly, and the necessary approximations lead to a regulator dependence of the endpoint, see also \fref{fig:FRG_regulator_theorySpace}.

As for equations of motion, one can derive equations for the correlation functions of a theory.
In this case, $n$ derivatives with respect to the fields are applied to \eref{eq:flow_eq} to obtain the flow equation of an $n$-point function.
The required differentiation rules are the same as for DSEs but with the propagator replaced by $\left(\Gamma_k^{(2)}[\Phi] + R_{k}\right)^{-1}$, see \fref{fig:fRG_master_prop_diffRules} and \eref{eq:DGammak} below.
Again, the external sources are set to zero at the end to obtain the propagators and vertices of the theory for fixed $k$.

As an example, consider the two-point function of a scalar theory.
Using the differentiation rules from \fref{fig:fRG_master_prop_diffRules}, the first derivative can only act on a propagator and yields (the two internal propagators are fully dressed)
\begin{align}\label{eq:FRGE-1deriv}
	\includegraphics[height=1.5cm]{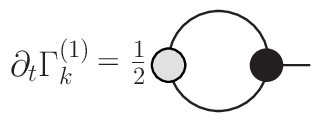}.
\end{align}
The second derivative can act on the two propagators or the vertex.
After setting the sources to zero, this leads to the flow equation for the two-point function:
\begin{align}\label{eq:FRGE-2p_fig} 
	\includegraphics[height=2cm]{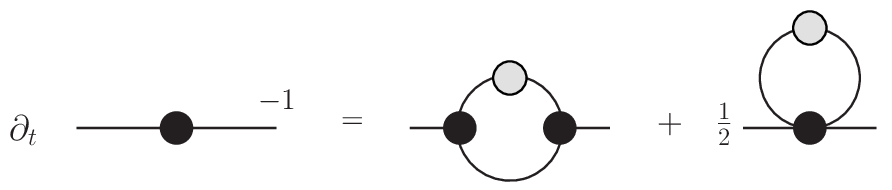}.
\end{align}
An algebraic derivation is provided in the box below.

It is clear from the structure of the Wetterich equation (\ref{eq:flow_eq}) and the differentiation rules in \eref{eq:derivatives-RG} that all flow equations only contain one-loop diagrams which is very convenient from a technical point of view.
Additionally, one can immediately see that in the flow equation for an $n$-point function only vertices up to order $n+2$ can appear.
In marked contrast to DSEs, only dressed vertices appear in flow equations.
As a consequence, additional topologies of diagrams can appear because no bare vertices are needed to construct them.
For example, to obtain the flow equation for the quark propagator in QCD, one simply replaces the external lines by quark lines in \eref{eq:FRGE-2p_fig} and then thinks of all possible (fully dressed) three- and four-point functions that can appear in the loop.
This leads to tadpole diagrams with quark-gluon, quark-quark and quark-ghost scattering kernels and a diagram with two quark-gluon vertices.
The quark propagator DSE, on the other hand, contains only a single loop diagram, as can be inferred from \fref{fig:dse_scalar_prop} using the same procedure.
The presence of the bare vertices, however, forbids the appearance of tadpole diagrams.
As one can see, although both equations describe the same quantity, they have a completely different structure.

A technical advantage of the FRG is that renormalization is implemented automatically via the regulator function which makes the integrals IR and UV finite.
Solving a flow equation requires calculating the integrals at fixed scale $k$ and solving the differential equation in $k$ with initial conditions given at $k_\text{in}$.
Lowering the scale $k$ can be interpreted as following an RG trajectory in theory space from $S[\Phi]$ to $\Gamma[\Phi]$.
As mentioned above, it is necessary to truncate the flow equations which leads in this picture to deviations from the exact trajectory, see \fref{fig:FRG_regulator_theorySpace}.

Before closing this section, let me mention a particularly attractive feature of the FRG, namely the possibility to
describe bound states and their constituents simultaneously in one framework using dynamical hadronization \cite{Gies:2001nw,Gies:2002hq,Pawlowski:2005xe,Floerchinger:2009uf,Floerchinger:2010da}; see \cite{Gies:2006wv} for a pedagogical introduction of the idea.\footnote{In the context of condensed matter and statistical physics systems it is called dynamical condensation/pairing.}
It is based on a Hubbard-Stratonovich transformation implemented in a scale-dependent way.
To this end, scale-dependent fields $\varphi$ for the bound states are introduced that carry the desired quantum numbers, for example,
\begin{align}
 \partial_t \varphi^a(p) = \int_q \partial_t A_{k,\overline{\psi}T\psi}(p-q,q) (\overline{\psi}T^a\psi )(p-q,q)
\end{align}
can be used to describe the $\pi$ and $\sigma$ mesons as bound states of quarks and antiquarks, $\psi$ and $\overline{\psi}$, respectively, when $T^a$ is chosen as generator of the flavor symmetry group.
$A_{k,\overline{\psi}T\psi}$ can be chosen such as to completely eliminate the corresponding channel of the four-Fermi interaction.
The scale dependence of the composite fields modifies the Wetterich equation:
\begin{align}
 \partial_t \Gamma_k[\Phi]=&\frac1{2} \text{Tr}\left\{ \left(\Gamma_k^{(2)}[\Phi] + R_{k}\right)^{-1} \partial_t R_{k}\right\}-\frac{\de \Gamma_k[\Phi]}{\de \varphi}\partial_t \varphi.
\end{align}

Dynamical hadronization provides a convenient way to include the physics of bound states in an efficient way.
It should be emphasized that dynamical hadronization is a way to include hadronic degrees of freedom explicitly.
In principle, the same can be achieved with the fundamental degrees of freedom, but with those the corresponding effects are encoded in higher order vertices which are difficult to include, see, e.g., \cite{Fu:2022uow,Fu:2024ysj,Fu:2025hcm} for corresponding FRG calculations in QCD.
Calculations including hadronic degrees of freedom via hadronization can be found, for instance in, \cite{Braun:2014ata,Mitter:2014wpa,Cyrol:2017ewj}.

\begin{BoxTypeA}[box:FRG]{Derivations: Wetterich equation and flow equation for the two-point function}

	Details of the derivations of the Wetterich equation (\ref{eq:flow_eq}) and the flow equation for the two-point function (\ref{eq:FRGE-2p_fig}) using the notation introduced in Sec.~\ref{sec:QFT} are provided here.

	The derivative of the effective average action with respect to $t=\ln(k/\Lambda)$ yields
	\begin{align}\label{eq:deriv-k}
	\partial_t \Gamma_k[\Phi]=& -\partial_t W_k[J]-\frac{\delta W_k[J]}{\delta J_i} \partial_t J_i+ (\partial_t J_i)\Phi_i- \frac1{2} \Phi_i (\partial_t R_{k,ij})\Phi_j\nnnl
	= & \left\langle \frac1{2} \phi_i (\partial_t R_{k,ij}) \phi_j \right \rangle_J -\frac1{2} \Phi_i (\partial_t R_{k,ij})\Phi_j
	= \frac1{2}  (\partial_t R_{k,ij}) \left( \left \langle \phi_i  \phi_j \right\rangle_J - \Phi_i \Phi_j \right)
	= \frac1{2}  (\partial_t R_{k,ij}) D_{k,ij}^{J}.
	\end{align}
	The second equation of \eref{eq:Gamma_k_Phi-k} was used to cancel the second and third terms in the first line.
	$\langle \phi_i \phi_j \rangle_J$ was decomposed as $D^{J}_{k,ij}+\langle \phi_i \rangle_J \langle \phi_j \rangle_J= D^{J}_{k,ij}+\Phi_i \Phi_j$ where $D^{J}_{k,ij}$ is the connected propagator in presence of the sources $J$, denoted by $\left(\Gamma_k^{(2)}[\Phi] + R_{k}\right)^{-1}$ above, at the regulator scale $k$:
	\begin{align}
	D^{J}_{k,ij}:=\frac{\delta^2 W_k[J]}{\delta J_i \delta J_j}.
	\end{align}
	Its inverse is the two-point function but with an additional contribution from the regulator $R_k$:
	\begin{align}\label{eq:DGammak}
	\de_{ij}=\frac{\de \Phi_i}{\de \Phi_j}=\frac{\de J_l}{\de \Phi_j} \frac{\de}{\de J_l} \frac{\de W_k[J]}{\de J_i}=
	\frac{\de \left(\Gamma_k[\Phi]+\frac{1}{2} \Phi_m R_{k,mn} \Phi_n \right)}{\de \Phi_j \de \Phi_l}  \frac{\de^2 W_k[J]}{\de J_l  \de J_i}=
	\left(\Gamma^{J}_{k,jl}[\Phi] + R_{k,jl} \right) D^{J}_{k,li}
	\end{align}
	with
	\begin{align}
	\Gamma^{J}_{k,ij}:=\frac{\de^2 \Gamma_k[\Phi]}{\de \Phi_i \de \Phi_j}, \qquad J_l=\frac{\de\left(\Gamma_k[\Phi]+\frac{1}{2}\Phi_m R_{k,mn} \Phi_n \right)}{\de \Phi_l}.
	\end{align}
	Eq.~(\ref{eq:deriv-k}) can thus be rewritten as
	\begin{align}\label{eq:flow_eq_indices}
	\partial_t \Gamma_k[\Phi]=& \frac1{2} \left(\Gamma^{J}_{k,ij}[\Phi] + R_{k,ij} \right)^{-1} \partial_t R_{k,ij},
	\end{align}
	which corresponds to the Wetterich equation (\ref{eq:flow_eq}) above.

	The differentiation rules shown in \fref{fig:fRG_master_prop_diffRules} are given by
	\begin{subequations}\label{eq:derivatives-RG}
	\begin{align}
	\frac{\de D_{k,ij}^{J}}{\de \Phi_l}&=\frac{\de\left(\Gamma^{J}_{k,ij} + R_{k,ij} \right)^{-1}}{\de \Phi_l}= D^{J}_{k,im} \Gamma^{J}_{k,mln} D^{J}_{k,nj},\\
	\frac{\delta}{\delta\Phi_{i}}\Gamma^{J}_{k,j_{1}\cdots j_{n}} & =\frac{\delta\Gamma_k[\Phi]}{\delta\Phi_{i}\delta\Phi_{j_{1}}\cdots\delta\Phi_{j_{n}}}=\Gamma^{J}_{k,ij_{1}\cdots j_{n}}.
	\end{align}
	\end{subequations}
	For signs see the comment below \eref{eq:deriv_D}.
	For Grassmann fields, the expected minus signs arise directly from their anticommutativity, see \cite{Huber:2019dkb} for details on realizations for corresponding algorithms.
	The algebraic derivation for the two-point function of a scalar theory can be done in one line:
	\begin{align}\label{eq:FRGE-2p}
	\partial_t& \Gamma^{J}_{k,ij}=\frac{\de^2}{\de \Phi_i \de\Phi_j} \partial_t \Gamma_k[\Phi]=\frac{1}{2} \frac{\de}{\de \Phi_i}  D^J_{mr} \Gamma^{J}_{k,rjs} D^J_{sn} \partial_t R_{k,mn} =
	D^J_{mt} \Gamma^{J}_{k,tiu} D^J_{ur} \Gamma^{J}_{k,rjs} D^J_{sn} \partial_t R_{k,mn} +
	\frac{1}{2}  D^J_{mr} \Gamma^{J}_{k,irjs} D^J_{sn} \partial_t R_{k,mn}.
	\end{align}
	This agrees with the graph in \eref{eq:FRGE-2p_fig}.

\end{BoxTypeA}

\subsection{Further functional equations}

The equations mentioned above are widely used in hadron physics.
Further types will only be mentioned briefly here and for more details I refer to the literature.

\begin{itemize}

\item Slavnov-Taylor identities encode the gauge symmetry on the level of correlation functions \cite{Taylor:1971ff,Slavnov:1972fg,Slavnov:2008sti}.
They do not form a complete set of equations but rather provide constraints on some parts of correlation functions.
For example, an STI fixes the longitudinal part of the gluon propagator in linear covariant gauges to be equal to the gauge fixing parameter $\xi$.
Their Abelian versions are called Ward-Takahashi identities \cite{Ward:1950xp,Takahashi:1957xn} which have a simpler structure due to the absence of ghosts.
In non-Abelian gauge theories, the so-called pinch technique can be used to obtain simpler ghost-free Ward identities, see \cite{Binosi:2009qm} for a detailed review.

\item A variational approach based on the Hamiltonian formulation of QCD was developed in Coulomb gauge \cite{Feuchter:2004mk}.
Using trial ans\"atze for the vacuum wave functional, its Schr\"odinger equation is solved by minimizing the energy.
This leads to equations that are similar to DSEs.
Using non-Gaussian ans\"atze provides access to vertices \cite{Campagnari:2010wc,Huber:2014isa}.
A covariant variational approach was developed in which the effective action in Landau gauge is minimized \cite{Quandt:2013wna}.

\item Nielsen identities were originally formulated for Abelian gauge theories \cite{Nielsen:1975fs} and describe the dependence on a gauge fixing parameter via a differential equation.
One application is to prove the gauge parameter independence of certain quantities like pole masses.
Starting from results in Landau gauge, $\xi=0$, Nielsen identities were used to extend results for propagators to linear covariant gauges, $\xi>0$ \cite{Breckenridge:1994gs,Aguilar:2015nqa,Napetschnig:2021ria}.

\item Correlation functions of composite operators can be written in a diagrammatic form.
This is based on the general identity \cite{Pawlowski:2005xe}
\begin{align}
\langle F(\phi) \rangle = F\left(\Phi+D^J\frac{\de}{\de \Phi}\right)
\end{align}
where $F(\phi)$ is an arbitrary functional of the fields.
For a composite operator $O(\phi(x))$, its propagator is then
\begin{align}
 \langle O(\phi(x)) O(\phi(y)) \rangle = O\left(\Phi(x)+D^J\frac{\de}{\de \Phi}\right) O\left(\Phi(y)+D^J\frac{\de}{\de \Phi}\right).
\end{align}
For $n$ fields in the expectation value, up to $n - 2$ loops can appear.
For a worked out example of the right-hand side see \cite{Huber:2019dkb} and for an application to transport coefficients \cite{Christiansen:2014ypa}.
\end{itemize}

\section{Working with functional equations}
\label{sec:workingWithFuncEqs}

The process of extracting information from functional equations consists of several steps.
Beyond their derivation, see Sec.~\ref{sec:funcEqs}, a truncation needs to be devised and a numerical framework has to be set up.
In many cases, analytic calculations in limiting cases provide additional insights.
Below, I discuss various aspects of this process including truncations, techniques, and available tools.

\subsection{Truncations}

Functional equations constitute infinitely large systems of coupled equations.
For most applications, in particular numerical calculations, this requires to restrict oneself to subsets of these systems which is called a truncation.
Beyond the literal meaning of the word in the sense of cutting some expansion at a certain point, this can also mean to use models and thereby decouple the equations.

In choosing a specific truncation, one has to consider two things: What does one need to calculate the target quantity with the desired accuracy, and what can one handle technically?
The former requires a solid understanding of the underlying mechanisms of and physical intuition for the problem.
To help with the second factor, external input or models can be helpful in achieving a good description of the target quantity.
Unfortunately, estimating the error of a truncation is in most cases quite challenging.
Nevertheless, it turns out that in those cases where this was tested, truncations seem to be well-behaved.
This is known as \textbf{apparent convergence} and means that beyond a certain point, adding further terms to a truncation has increasingly smaller effects on the results, see \cite{Ihssen:2024miv,Huber:2025kwy} and references therein for dedicated calculations.
Since testing a truncation by enlarging it is quite cumbersome and not always possible for purely practical reasons, an alternative way to assess its reliability is the comparison of intermediate results with those from other methods.
For example, propagator dressings can be compared to lattice results and these propagators can be used to calculate bound states.

In the following, I discuss some common truncations.
For methods based on homotopy analysis and Monte Carlo sampling \cite{Pfeffer:2018tkx} or the Hopf algebra \cite{Kreimer:2006ua}, I refer to the literature.

\subsubsection{Vertex expansion}
\label{sec:vertexExpansion}

This is the standard truncation for DSEs and can also be used for flow equations.
As the 1PI effective action is the generating functional for (1PI) correlation functions, see \eref{eq:Gamma_expansion}, a direct approach to cut the infinite tower of equations is to truncate this series and take into account only a certain number of vertices.
I discuss such a truncation for the 1PI effective action and comment on the realization for the effective average action at the end.

DSEs can be derived as exact equations.
The necessity of truncations becomes obvious as the DSE for an $n$-point function contains also $m$-point functions with $m>n$.\footnote{For a theory which contains as highest bare vertex one with $l$ legs, $m$ can be at most $n+l-2$. The highest interaction also determines the maximal number of loops in a DSE as $l-2$.}
As an example see \fref{fig:dse_scalar_prop} where the DSE for the propagator of the scalar theory in \eref{eq:scalarLagrangian} is shown.
The propagator DSE contains the fully dressed three- and four-point functions which have their own DSEs.
To solve a specific DSE, one has to specify how to treat all the other correlation functions on which it depends.
The simplest possibility to treat such an external correlation function is to set it to zero, thereby literally truncating the effective action.
Alternatively, one can use a given expression for it, be it a model, a result from another method, or the solution of its own DSE.

The use of models thereby offers a great flexibility as the employed model does not necessarily have to be a good approximation of the true form of the correlation function it represents.
Hence it can be used to counteract the effect of other discarded terms.
A prime example of this is the so-called rainbow truncation of the quark propagator DSE in which only one tensor of the quark-gluon vertex is included and modelled together with the gluon propagator by a simple function.
Together with the ladder truncation for hadronic bound state equations, this provides an efficient truncation for many applications in hadron physics, see \cite{Eichmann:2025wgs} for more details.

For the construction of a truncation several guides can be used.
One is a consistent treatment of the perturbative regime by including the corresponding number of loops.
While this reproduces the correct behavior at a certain order of the coupling, it is also possible to realize the resummed perturbative behavior in form of the anomalous dimension either be adapting the models accordingly or by including the terms of the resummation explicitly, see \cite{Huber:2018ned}.
Another important guide is symmetries.
The most relevant example here is certainly chiral symmetry in the quark sector which leads to the axial-vector Ward-Takahashi identity that relates the bound state kernel of a meson to the quark selfenergy, see \cite{Eichmann:2025wgs} for details.
In this case, a symmetry violating truncation directly affects the pion which looses its status as a Goldstone boson.\footnote{Since quarks are not massless in nature, chiral symmetry is not an exact symmetry of QCD but a good approximate one and the related Goldstone bosons are not massless but light.
However, calculating with massless quarks and testing the Goldstone nature is a relatively simple and important test.}
A third guide is the quantitative importance of a correlation function which can be estimated or explicitly tested by exploratory calculations.

In a similar way, the effective average action $\Gamma_k[\Phi]$, being the generating functional for $k$-dependent correlation functions, can be truncated.
Deriving flow equations for the $n$-point functions was already discussed in Sec.~\ref{sec:FRG}.
The mechanism of truncating the infinite tower of equations is then completely analogue to the DSE case.
The vertex expansion is typically used when the full momentum dependences of the correlation functions are required which are not accessible with the derivative expansion discussed in Sec.~\ref{sec:derivExp}.

\subsubsection{Loop expansion}
\label{sec:loopExp}

Although the equations of motion from $n$PI effective actions look deceptively similar to DSEs, there are important differences.
In contradistinction to DSEs, the number of equations of motion is finite for $n$PI effective actions, simply because there is only a finite number of correlation functions included.
To write down these equations, the $n$PI effective action is typically expanded in terms of nonperturbative loops, see Figs.~\ref{fig:Gamma2} and \ref{fig:3PI_3l_YM} for examples.
This leads to expressions similar to DSEs.
In some cases, they even agree with them.
Explicitly worked out examples for this include the scalar (as well as the gluon) propagator for the 4PI effective action at four-loops \cite{Carrington:2009kh}.
Due to interacting only via three-point functions, the ghost and quark propagator equations of motion are exact for the 3PI effective action at three loops \cite{Berges:2004pu}.
However, the equivalence between the equations of motion of the 1PI and $n$PI effective actions is only formal as there is an implicit truncation dependence via other (higher) correlation functions.

It should be appreciated that for $n$PI effective actions vertices with more than $n$ legs are not included dynamically but their effects are encoded in the higher loops.
Thus, using an $n$PI effective action does \textit{not} mean to truncate in the vertices as opposed to the vertex expansion and the resulting equations, despite the similarities to DSEs, can be seen as complementary to DSEs.
As an example, consider the three-gluon vertex for which results from DSEs and the 3PI effective action truncated to three loops exist \cite{Huber:2020keu}.
The 3PI equation of motion is a one-loop equation which captures the main features of the vertex quite well.
The DSE, on the other hand, is a two-loop equation that leads to results that agree very well with the 3PI results.
However, despite the similarities at the one-loop level, the two-loop contributions are important in the latter \cite{Huber:2020keu}.

More details on truncations of $n$PI effective actions are discussed in Sec.~\ref{sec:glueballSpectrum} with an explicit example.

\subsubsection{Derivative expansion}
\label{sec:derivExp}

An expansion in derivatives is a widely used truncation of flow equations.
As suggested by the name, the scale-dependent effective action $\Gamma_k[\Phi]$ is expanded in derivatives which corresponds to an expansion in momenta of the vertices.
For illustration purposes, let us consider an $N$-component field with $O(N)$ symmetry.
This model can already demonstrate basic concepts of symmetry breaking and restoration and describes, depending on the number of dimensions $d$ and the value for $N$, different physical applications, see, e.g., \cite{Berges:2000ew} for an overview.
Due to the symmetry, the ansatz for the effective action can only contain appropriate combinations of the fields, and we will use the convenient notation $\rho=\Phi^2/2$.
To second order, the derivative expansion reads
\begin{align}\label{eq:ON_deriv_exp}
	\Gamma_k[\Phi] = \int d^dx \left(
	 \frac{Z_k(\rho)}{2} (\partial_\mu \Phi)^2 + \frac{Y_k(\rho)}{4} (\partial_\mu \rho)^2 + V_k(\rho)
	 + \mathcal{O}(\partial^4)
	\right)
\end{align}
with $Z_k=1$ and $Y_k=0$ at the initial RG scale.
The lowest order is known as local potential approximation (LPA).
It contains only the kinetic term but with $Z_k=1$ and the potential $V_k(\rho)$:
\begin{align}\label{eq:ON_LPA}
	\Gamma_k^\text{LPA}[\Phi] = \int d^dx \left(
		\frac{1}{2} (\partial_\mu \Phi)^2 + V_k(\rho)
		\right).
\end{align}
From this ansatz, the flow equation for the effective average potential $V_k$ can be derived.
By expanding it in a Taylor series around the minimum at $\rho_0$, one obtains flow equations for infinitely many couplings.

By definition, the field has no anomalous dimension $\eta=-k\,\partial  \ln Z_k/\partial k$ within the LPA.
To overcome that, the wave function renormalizations $Z_k$ can be included.
This is often called LPA' ("LPA prime"):
\begin{align}\label{eq:ON_LPAp}
	\Gamma_k^\text{LPA'}[\Phi] = \int d^dx \left(
	 \frac{Z_k(\rho)}{2} (\partial_\mu \Phi)^2 + V_		k(\rho)
	\right).
\end{align}
For the complete second order in the derivative expansion shown in \eref{eq:ON_deriv_exp}, terms with two derivatives of invariant field combinations need to be included.
While higher orders have been considered for the $O(N)$ model \cite{Balog:2019rrg,DePolsi:2020pjk}, it becomes very cumbersome for other cases.

One limitation of the derivative expansion is that the full momentum dependence of correlation functions is not captured.
If this is required, the vertex expansion discussed in Sec~\ref{sec:vertexExpansion} is the method of choice.

\subsection{Techniques}

Solving functional equations ranges from simple cases that can be done by hand to computationally demanding cases requiring computer clusters.
Here I discuss techniques which are relevant for the topic discussed in Sec.~\ref{sec:glueballSpectrum}, the calculation of the glueball spectrum using bound state equations and equations of motion.
Nevertheless, some of them are also useful for flow equations, in particular when working with a vertex expansion.
For more details on techniques like an expansion of the effective average potential in field space \cite{Berges:2000ew}, grid methods \cite{Bohr:2000gp}, pseudospectral methods \cite{Borchardt:2016pif} or finite element methods \cite{Grossi:2019urj,Sattler:2024ozv} I refer to the literature.

\subsubsection{Solving equations of motion}

Once a truncation has been devised, the resulting functional equations need to be solved numerically.
Only in limiting cases it is possible to obtain analytic solutions.
This includes the perturbative regime where beyond the fixed order also solutions taking into account the anomalous runnings of dressing functions can be obtained using an approximation for the angular integral, see, e.g., \cite{vonSmekal:1997vx,Fischer:2003zc,Huber:2018ned}.
Also in the IR limit the qualitative behavior can be extracted.
In fact, this is one of the rare cases for which a (qualitative) solution for the complete set of equations, viz., without truncations, can be found \cite{Alkofer:2004it,Alkofer:2008jy,Fischer:2009tn,Huber:2009wh}.

For quantitative solutions, it is necessary to solve the equations numerically.
To express the dressing functions over a finite range of momenta, standard numerical techniques can be used.
The most common approach is interpolation on a grid of the kinematic variables.
Alternatively, they can be expanded in a basis of functions, for example, Chebyshev polynomials.
The calculation of the integrals can usually be done with standard Gaussian quadrature methods.
For massless propagators some care needs to be taken to avoid numerical instabilities in the integrand, though, see, e.g., \cite{Huber:2012kd,Huber:2011xc}.
Two methods have become standard for solving the equations.
The more common one, fixed-point iteration, is also the simpler one.
Starting from an ansatz for a correlation function, the right-hand side is evaluated to determine a new expression for the correlation function.
The new expression is used to calculate the right-hand side again.
This is repeated until the changes are below the desired accuracy.
Iteration works well in many cases.
Sometimes it is necessary to introduce a relaxation parameter, viz., the old and new solutions are combined to dampen oscillations of the solution.
When several coupled equations need to be solved, the principle is the same but require some exploration of the best order of iteration and the introduction of meta-iterations (groups of iterations).

If simple iteration fails or is not stable enough, minimization may be a viable alternative \cite{Atkinson:1997tu}.
Thereby, the equations are rewritten into the following form:
\begin{align}
 E = -D+D_0+\Sigma,
\end{align}
where a simplified symbolic notation suppressing all momentum dependences is used.
$D$ is the full dressing function, $D_0$ is the bare term and $\Sigma$ is the sum of diagrams.
$E$ is the quantity to be minimized: only for $E=0$ we have a solution.
The minimization acts upon the expansion coefficients of the dressing functions.
A common choice is the Newton-Raphson method to find the minimum.
Denoting the expansion coefficients by $c_i$, one step of the minimization procedure to calculate new expansion coefficients $c_i'$ is realized by
\begin{align}
 c_i' = c_i - \lambda \sum_j J_{ij}^{-1} E_j,
\end{align}
where $\lambda$ is a relaxation parameter and $J_{ij}$ the Jacobian given by
\begin{align}
 J_{ij} = \frac{\partial E_i}{\partial c_j}.
\end{align}
Instead of calculating $J$ exactly, it is sufficient to approximate it with Broyden's method using a simple forward differentiation:
\begin{align}\label{eq:Broyden}
 J_{ij} \approx \frac{E_i(c_j+h)-E_i(c_j)}{h}.
\end{align}
Typical values for $h$ and also $\lambda$ can be found in \cite{Huber:2012cd}.
While this method tends to be more stable than simple iteration, it is slower due to the calculation of the Jacobian.
For $N$ momentum configurations, the integrals need to be evaluated $N$ times for standard iteration and $N^2$ times for the minimization method (times two if \eref{eq:Broyden} is used).
In the context of the FRG, Newton-Raphson iteration appears with pseudospectral methods \cite{Borchardt:2016pif}.

A third solution technique is based on spectral representations.
This includes the Nakanishi integral representation, see, e.g., \cite{Sauli:2001mb,Jia:2017niz,Mezrag:2020iuo}, or spectral DSEs, see, for instance, \cite{Horak:2020eng,Horak:2022aza}.
Neural networks are only beginning to be explored \cite{Terin:2024iyy}.

Before the numerical methods described above can be applied, there is one more thing to consider: Most integrals are divergent.
This has its roots already in perturbation theory.
The proper treatment of these divergences is called renormalization, see \cite{Pascual:1984zb} for an exhaustive explanation for QCD including detailed calculations.

\begin{BoxTypeA}[box:renorm]{Renormalization in a nutshell}
	
Renormalization consists of the process of regularizing the divergent integrals and removing them by renormalization prescriptions.
When working with perturbation theory, dimensional regularization is often favored for the former step.
In numerical calculations, though, regularization is most conveniently realized by a UV momentum cutoff $\Lambda$ of the integral.
The dependence on the cutoff $\Lambda$ can be determined analytically due to the perturbative origin of the divergences.
This knowledge can be useful for numeric calculations.

Let us consider a generic equation of motion of the following form where the divergences are logarithmic in the cutoff:
\begin{align}\label{eq:genericEoM}
 D(p;\mu) = Z(\mu,\Lambda)  + \Sigma(p; \Lambda).
\end{align}
This could be, for example, the DSE for the quark mass function, $D(p;\mu)=M(p)$.
The dependence of the selfenergy $\Sigma(p;\Lambda)$ on the cutoff $\Lambda$ is explicitly indicated and has the form $\ln(\Lambda/p)$.
$Z(\mu,\Lambda)$ is a renormalization constant which comes from the bare dressing function and is introduced in the renormalization process to subtract the divergences.
Thus, it must depend on the cutoff in such a way that the left-hand side is independent of $\Lambda$.
To determine $Z(\mu,\Lambda)$, one fixes the dressing at a given renormalization point $\mu$:
\begin{align}
 Z(\mu,\Lambda) = D(\mu;\mu) - \Sigma(\mu;\Lambda).
\end{align}
Obviously, the renormalization prescription is not unique and different choices lead to different renormalization constants, depending on the choices of the renormalization point $\mu$ and the value of $D(\mu;\mu)$.
As a consequence, also the result for $D(p;\mu)$ depends on the renormalization prescription:
\begin{align}\label{eq:renormEoM}
 D(p;\mu) = D(\mu;\mu) + \Sigma(p;\Lambda) - \Sigma(\mu;\Lambda).
\end{align}
The logarithmic divergences cancel out as $\ln(\Lambda/p)-\ln(\Lambda/\mu)=\ln(\mu/p)$.
In this way, one trades the dependence on the cutoff $\Lambda$ for a dependence on the renormalization point $\mu$.
While it seems arbitrary that we introduced this dependence on $\mu$, it is actually a feature of the renormalization procedure and leads to the concept of the renormalization group.
Thus, correlation functions do depend on the renormalization point and the renormalization prescription, but they transform according to the rules of the renormalization group under changes of these parameters.
In physical observables, however, this dependence cancels out.
The attentive reader might have noticed that the quark mass function $M(p)$ mentioned above does not depend on the renormalization point.
This is an example of a renormalization group invariant quantity (but not a physical observable in this case).
The invariance comes about from the definition of $M(p)$ as $B(p;\mu)/A(p;\mu)$, where the $\mu$-dependences of the quark propagator dressing functions $A$ and $B$ cancel out.
The process of renormalizing as in \eref{eq:renormEoM} is called a MOM scheme, where MOM stands for momentum subtraction.

Eq.~(\ref{eq:genericEoM}) describes the basic idea of renormalization as also used in perturbative calculations.
Additional complications arise if the selfenergy includes further renormalization constants, for example for each bare vertex in the case of DSEs.
Depending on the system one solves, the value of these renormalization constants can be determined from their own equations or, in gauge theories like QCD, by relations between renormalization constants called Slavnov-Taylor identities which encode gauge symmetry.
A consistent implementation of renormalization for larger systems turns out to be non-trivial, see \cite{Huber:2020keu,Eichmann:2021zuv} for examples, but is crucial both for respecting gauge symmetry and for the resummation of diagrams to obtain the correct anomalous dimensions \cite{Huber:2018ned,Huber:2020keu}.
A special treatment is also required if quadratic divergences appear, see, for example, \cite{Huber:2014tva,Huber:2020keu,Eichmann:2021zuv}

\end{BoxTypeA}

\subsubsection{Solving bound state equations}
\label{sec:solveBSEs}

\begin{figure}[tb]
	\raisebox{1cm}{\includegraphics[width=0.45\textwidth]{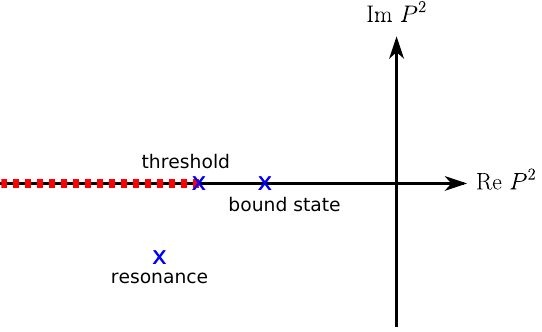}}\hfill
	\includegraphics[width=0.35\textwidth]{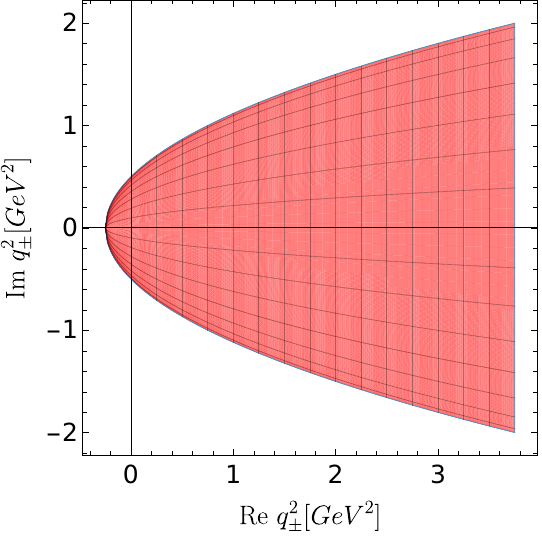}
	\caption{Left: Poles on the real line with $P^2<0$ correspond to bound states, poles on the second Riemann sheet to resonances.
	Branch cuts start at thresholds.}
	Right: Region of integration in the complex $q_{\pm}^2$ plane as determined from \eref{eq:parabola} with $M^2=1\,\text{GeV}$.
\label{fig:parabola}
\end{figure}

The standard method to solve a (homogeneous) bound state equation is to rewrite it into an eigenvalue problem by introducing a parameter~$\lambda$:
\begin{align}\label{eq:BSE2}
	\int_q K(p,q,P)\,G_0(q,P)\,\Gamma(q,P) = \lambda(P^2)\,\Gamma(p,P),
\end{align}
or, in a more compact notation,
\begin{align}
	K\,G_0\,\Gamma = \lambda\,\Gamma.
\end{align}
For simplicity, we consider two-body bound states, but the method can be generalized to more complicated systems.
After discretizing the integral (with integration weights implicitly contained in the sum),
\begin{align}
 \sum_j \left[K(p_i,q_j,P)\,G_0(q_j,P)\right]\,\Gamma(q_j,P) = \lambda(P^2)\,\Gamma(p_i,P),
\end{align}
the eigenvalue problem for the matrix $[K\,G_0]$ can be solved for a given value of $P^2$.
This value is then varied until an eigenvalue $\lambda=1$ is found which corresponds to a bound state of mass $M^2=-P^2$.
The ground state, being the lightest state, corresponds to the highest eigenvalue curve which hits 1 at the highest $P^2$ (thus, lowest $M$) and lower eigenvalue curves correspond to excited states.
Also resonances are described by bound state equations, see \fref{fig:parabola}.
In that case, eigenvalues are calculated on the first Riemann sheet and continued analytically to the second Riemann sheet to find a solution for complex $P^2$, see \cite{Santowsky2020} for an example.
The four-dimensional integration over $q$ is split as follows.
One angular integral can be done analytically, one angular integral is done numerically, and the other two integrals, the radial part and the remaining angular one, are the discretized ones.
They correspond to the columns of the matrix $[K\,G_0]$, while the radial part and one angle of the external momentum $p$ correspond to the rows.

Eq.~(\ref{eq:BSE2}) is a homogeneous integral equation.
Consequently, the resulting Bethe-Salpeter amplitudes in form of the eigenvectors are not normalized.
For many applications, however, like decay constants or form factors, normalization is required.
There exist several ways to achieve this, for example, by considering the next order of the expansion of the scattering matrix around the pole or from Ward-Takahashi identities, see \cite{Cutkosky:1964zz,Nakanishi:1965zza,Nakanishi:1969ph,Predazzi:1965bse,Eichmann:2016yit,Sanchis-Alepuz:2017jjd} for more details.

One complication of calculating the matrix comes from the propagator $G_0$.
The arguments of the individual propagators contain the total momentum $P$ of the bound state which is necessarily imaginary for a solution due to $M^2=-P^2$: $P=(0,0,0,i\,M)^T$.
For a two-body bound state the arguments of the propagators can be parametrized as $q_{\pm} = \frac{1}{2}P\pm q
$ and hence the propagators are probed at
\begin{align}\label{eq:parabola}
 q_{\pm}^2 = \frac{1}{4}P^2\pm q\cdot P + q^2 = -\frac{1}{4}M^2\pm i\,|q|\,M\cos\theta_q + q^2.
\end{align}
The boundaries with $\cos\theta_q=\pm1$ correspond to a parabola in the complex plane with the apex at $-M^2/4$, see \fref{fig:parabola}.
The calculation of correlation functions in the complex plane comes with its own challenges which are discussed in the next section.
Depending on the truncation of the kernel, also the correlation functions appearing in it need to be known for complex arguments.

\subsubsection{Calculations for complex momenta}
\label{sec:complexMomenta}

As explained above, we need the correlation functions appearing in bound state equations in the complex momentum plane and know their analytic structure.
Another case where this knowledge is required is the calculation of transport coefficients.
Unfortunately, obtaining such solutions is not as simple as replacing the real momenta by complex ones in the integrals.
The reason is that there are branch cuts and poles in the integrands which have to be taken into account properly, see \cite{Windisch:2013dxa,Huber:2023uzd} for an introduction to the problem for two- and three-point functions.

There are several methods to solve the equations for complex external momenta.
Direct calculations can be done if the integration contour is deformed appropriately to avoid the branch cuts and poles \cite{Maris:1995ns,Alkofer:2003jj,Eichmann:2007nn,Windisch:2012sz,Eichmann:2019dts,Fischer:2020xnb,Huber:2022nzs}.
The complication of this approach is that the position of the branch cuts depend on the external momenta and needs to be known beforehand to determine a safe integration contour.
The shell method \cite{Fischer:2005en,Fischer:2008sp} and the Cauchy method \cite{Krassnigg:2009gd} can be used to solve the equations for momenta in the region required for the bound state equations, the parabola defined in \eref{eq:parabola}.
Complementary methods rely on spectral representations, e.g., \cite{Jia:2017niz,Mezrag:2020iuo,Horak:2022myj,Duarte:2022yur,Braun:2022mgx,Eichmann:2023tjk,Horak:2023hkp,Pawlowski:2024kxc,Pawlowski:2025etp}.
For example, a propagator can be expressed in terms of a spectral density $\rho(s)$ as:
\begin{align}\label{eq:spectRepr}
 D(p^2) = \int_{s_0}^\infty ds \frac{\rho(s)}{p^2+s}.
\end{align}
$s_0$ is a threshold value below which the spectral density vanishes.
If there are poles on the real line, they can be included in $\rho(s)$ via Dirac delta distributions.
Many calculations lead to poles in the complex plane which would lead to an extra contribution in \eref{eq:spectRepr}.
For propagators the spectral representation in the form of \eref{eq:spectRepr} is physically related to the K\"all\'en-Lehmann representation.
Mathematically, it is a direct consequence of Cauchy's integral formula.
Finally, indirect methods are also used which use the results from real momenta to numerically reconstruct dressing functions for complex momenta.
To name  a few possibilities, this can be fits with trial functions, the Bayesian spectral reconstruction method, machine learning methods, Gaussian processes, the Tikhonov regularization, or Pad\'e approximants in various forms; see, for
instance, \cite{Cyrol:2018xeq,Dudal:2019gvn,Binosi:2019ecz,Li:2019hyv,Horak:2021syv,Lechien:2022ieg,Falcao:2022gxt,Boito:2022rad}.

\subsection{Tools}

\begin{table}[t]
	\TBL{\caption{List of publicly available programs and tools useful for working with functional equations.
	\label{tab:programs}}}{
		\begin{tabular*}{\columnwidth}{@{\extracolsep\fill}p{2.9cm}lp{5cm}l}
			\toprule
			\multicolumn{1}{@{}l}{\TCH{Program}} &
			\multicolumn{1}{@{}l}{\TCH{Current version}} &
			\multicolumn{1}{@{}l}{\TCH{Purpose}} &
			\multicolumn{1}{@{}l}{\TCH{Notes}} \\
			\colrule
			\textit{DiFfRG} \cite{Sattler:2024ozv} & & Solving FRG equations & \textit{C++}\\
			\textit{DoFun} \cite{Huber:2019dkb} & 3.0.1 & Derivation of DSEs, FRG equations, correlation functions of composite operators & \textit{Mathematica} package\\
			\textit{FeynCalc} \cite{Shtabovenko:2023idz} & 10.1.0 & Working with quantum field theoretical expressions and calculating Feynman diagrams & \textit{Mathematica} package\\
			\textit{FORM} \cite{Ruijl:2017dtg,FORM:2026frm} & 5.0.0  & Tracing indices, simplification and optimization & \\
			\textit{FormTracer} \cite{Cyrol:2016zqb} & 2.3.6 & Tracing indices & \textit{Mathematica} package using \textit{FORM}\\
			\textit{QMeS} \cite{Pawlowski:2021tkk} & 1.2 & Derivation of DSEs, FRG equations, STIs & \textit{Mathematica} package\\
			\textit{TensorBases} \cite{Braun:2025gvq} & - & Tensor bases & \textit{Mathematica} package\\
			\textit{Mathematica} \cite{Mathematica:2025url} & 14.3 & Tracing indices, simplification and optimization & \\
			\colrule
			\textit{CrasyDSE} \cite{Huber:2020ymd} & 1.1.0 & Solving DSEs numerically & \textit{C++}, no longer maintained\\
			\botrule
	\end{tabular*}}{}
\end{table}

The derivation of functional equations and solving them can easily become quite cumbersome due to the number of terms and their complexity.
Computer programs can thus be of great help and have been developed for the different steps.
This section provides a short overview of the main available programs and their capabilities.

For the derivation of various functional equations two dedicated programs are available publicly in form of \textit{Mathematica} packages.
With \textit{DoFun} (Derivation of functional equations) \cite{Alkofer:2008nt,Huber:2011qr,Huber:2019dkb} and \textit{QMeS} (Quantum Master equations:
environment for numerical Solutions) \cite{Pawlowski:2021tkk} Dyson-Schwinger equations, functional renormalization group equations, correlations functions of composite operators and Slavnov-Taylor identities can be derived, see \tref{tab:programs} for details.
The construction of tensor bases can be facilitated with \textit{TensorBases} \cite{Braun:2025gvq}.
The output of these programs can be further processed using tools like \textit{FORM} \cite{Ruijl:2017dtg}, \textit{Mathematica} \cite{Mathematica:2025url}, \textit{FormTracer} \cite{Cyrol:2016zqb} or \textit{FeynCalc} \cite{Mertig:1990an,Shtabovenko:2016sxi,Shtabovenko:2020gxv}.
For solving the equations numerically, \textit{C}, \textit{C++}, \textit{Fortran}, \textit{Python}, \textit{Julia}, or, in simple cases, \textit{Mathematica} are typically used.
For beginners, the following references and repositories might be helpful.
Ref.~\cite{Maas:2005xh} discusses numerical techniques specifically for solving the two Yang-Mills propagators.
\textit{CrasyDSE} \cite{Huber:2011xc} is a \textit{C++} program with the possibility to write integration kernels in \textit{Mathematica}.
Although it is no longer maintained, it might still offer some useful guidance for beginners.
\textit{DiFfRG} offers a framework for solving functional renormalization group equations \cite{Sattler:2024ozv} in \textit{C++}.
Finally, a DSE code written in \textit{Julia} is also available \cite{Zierler:2023qvz,Zierler:2024cod}.

\section{Functional methods at work: From correlation functions to the glueball spectrum}
\label{sec:glueballSpectrum}

In this section I illustrate the application of different functional equations to obtain the spectrum of pure Yang-Mills theory.
After discussing the setup, the calculation of the required correlation functions from equations of motion and their usage in bound state equations will be described.
The results in this section were obtained following a top-down approach where all relevant correlation functions are calculated.
For an example using the bottom-down approach, which relies on phenomenologically motivated models, see \cite{Eichmann:2025wgs}.

\subsection{Three-loop truncation of the 3PI effective action}

\begin{figure}[b]
	\centering
		\includegraphics[width=0.8\textwidth]{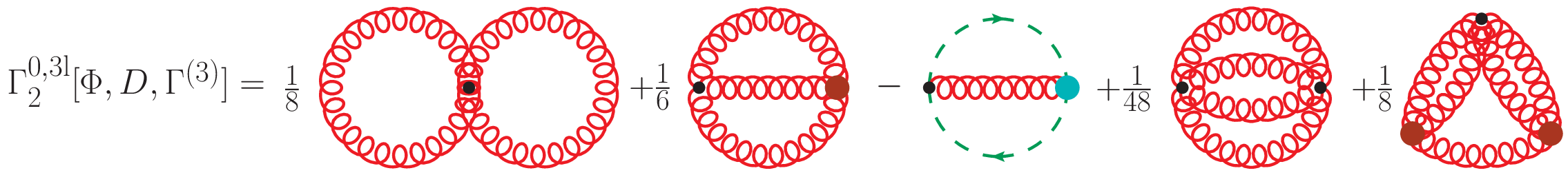}\\
		\vskip2mm
		\includegraphics[width=0.8\textwidth]{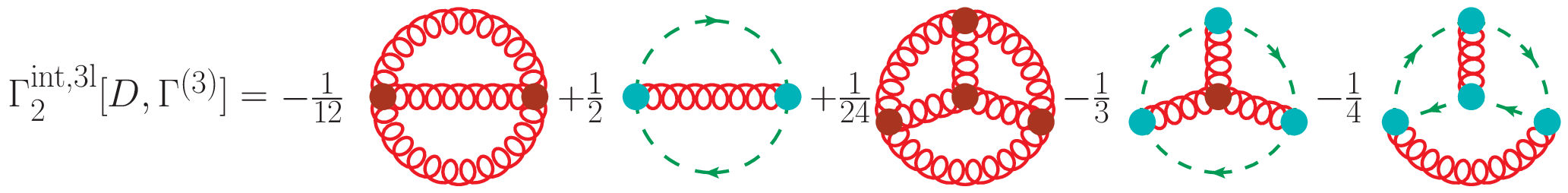}
		\caption{Contributions to the three-loop 3PI effective action for pure Yang-Mills theory, $\Gamma_2^\text{3l}=\Gamma_2^\text{0,3l}+\Gamma_2^\text{int,3l}$.
		Wiggly lines denote gluon propagators and dashed ones ghost propagators.
		Dots represent bare and disks dressed vertices.}
		\label{fig:3PI_3l_YM}
\end{figure}

In the absence of any obvious guide like chiral symmetry in the quark sector, the least we can demand for pure Yang-Mills theory is a setup that offers a consistent truncation of both the equations of motion of the correlation functions and the bound state equations.
The 3PI effective action and its truncation at three loops, shown in \fref{fig:3PI_3l_YM}, offers that as it includes the three-gluon vertex, which encodes the driving interaction between gluons to form a glueball, as a dynamic quantity.
Working in Landau gauge, we also have to deal with ghost fields from the Faddeev-Popov procedure.
They are scalar but anticommuting fields.
The form of the truncated action can be directly inferred from our discussion in Sec.~\ref{sec:nPI}.
For the ghost fields, closed loops acquire a minus sign and the symmetry factors need to be adapted.

The 3PI effective action offers several advantages.
First, it allows a systematic truncation in terms of loops \emph{at the level of the action}.
The resulting system of equations is fully self-contained and does not have any free parameters if all equations are solved.
Another one is that for the three-gluon vertex satisfactory results are obtained already for the three-loop truncation which results in a one-loop equation of motion.
To reach the same level of precision with a DSE, also its two-loop diagrams are required which are computationally more demanding \cite{Huber:2020keu}.

\begin{figure}
		\includegraphics[width=0.98\textwidth]{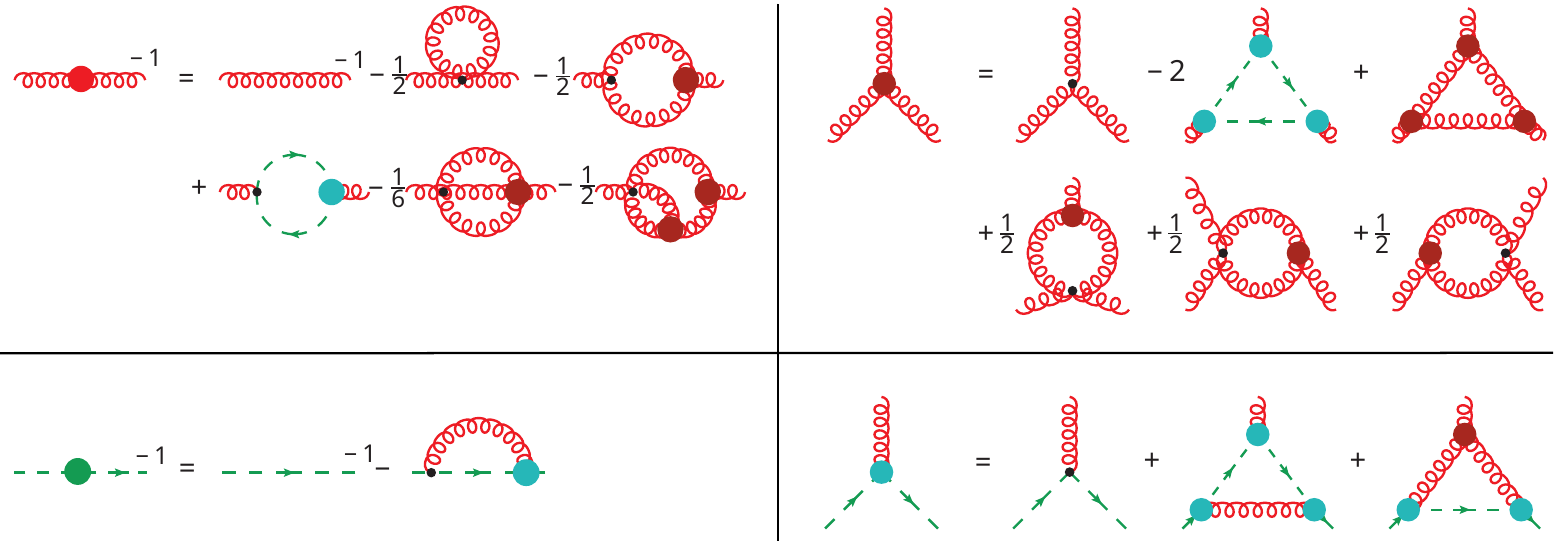}
		\caption{The equations of motion for the gluon propagator (top left), the ghost propagator (bottom left), the three-gluon vertex (top right) and the ghost-gluon vertex (bottom right) for pure Yang-Mills theory.
		The two equations on the left are exact DSEs.
		Upon replacing the dressed for-gluon vertex by the bare one, the four equations are the equations of motion of the 3PI effective action at three loops.
		}
	\label{fig:3PI_3l_YM_EOM}
\end{figure}

Following the procedure outlined in Sec.~\ref{sec:nPI} the equations of motion for the propagators and the three-point functions can be derived.
They are shown in \fref{fig:3PI_3l_YM_EOM} with one modification for the gluon propagator which is promoted to the exact DSE by inserting a dressed four-gluon vertex into the sunset diagram.
This is advantageous for recovering the correct anomalous dimension of the gluon propagator.
The four-gluon vertex itself is calculated from its DSE, see \cite{Huber:2020keu}.

\begin{figure}[t]
	\centering
	\includegraphics[width=0.7\textwidth]{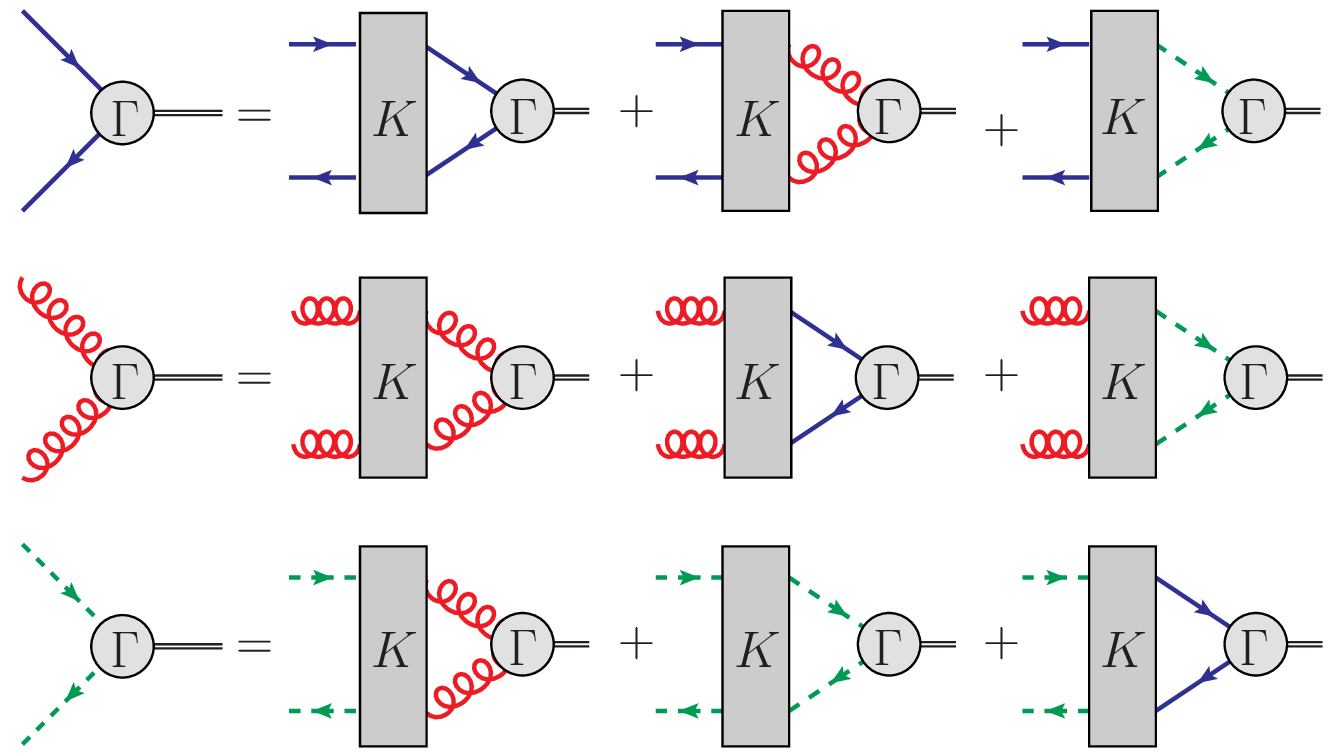}
	\caption{The glueball BSE for full QCD.
	The blue continuous lines with arrows denote quark, the red wiggly ones gluon and the green dashed ones ghost propagators.
	The gray boxes represent the different scattering kernels, for example, the quark-quark, quark-gluon, and quark-ghost scattering kernels in the first line.
	The gray disks are the Bethe-Salpeter amplitudes.
	For pure Yang-Mills theory, the quark-antiquark Bethe-Salpeter amplitudes can simply be set to zero.
	}
	\label{fig:BSE_glueball_YM}
\end{figure}

\begin{figure}[t]
	\centering
	\includegraphics[width=0.95\textwidth]{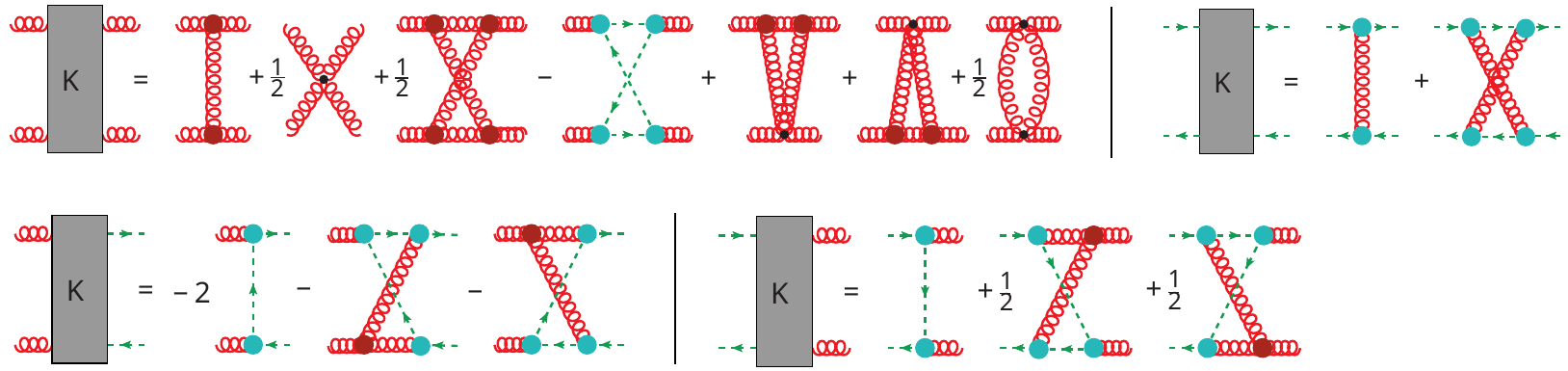}
	\caption{The BSE kernels of Yang-Mills theory from a three-loop truncated 3PI effective action.
	}
	\label{fig:kernels_YM_3PI}
\end{figure}

The kernels for the BSE, shown in \fref{fig:BSE_glueball_YM} for full QCD, are derived using \eref{eq:kernel}.
They can also be inferred directly from \eref{eq:3PI-kernel}.
As we have two types of fields, there is also a scattering kernel between ghosts and gluons which couples a ghost amplitude to a gluon amplitude.
Consequently, we need to include the BSE for two ghosts as well.\footnote{When calculating mesons, there is also a mixed kernel between quarks and gluons, which, however, is not relevant as long as there is no mixing with the corresponding gluonic states.}
This does not lead to an additional state in the spectrum but is merely reflecting the fact that we are working in Landau gauge.
The four kernels resulting from the action in \fref{fig:3PI_3l_YM} are depicted in \fref{fig:kernels_YM_3PI}.
A direct consequence of using a three-loop truncation of the effective action is the appearance of one-loop diagrams in the kernels and hence two-loop diagrams in the BSEs.

\subsection{Correlation functions}

\begin{figure}[t]
	\centering
	\includegraphics[width=0.48\textwidth]{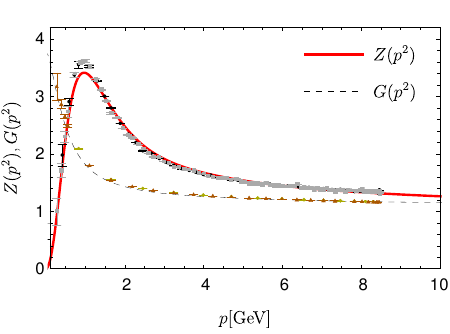}\hfill
	\includegraphics[width=0.48\textwidth]{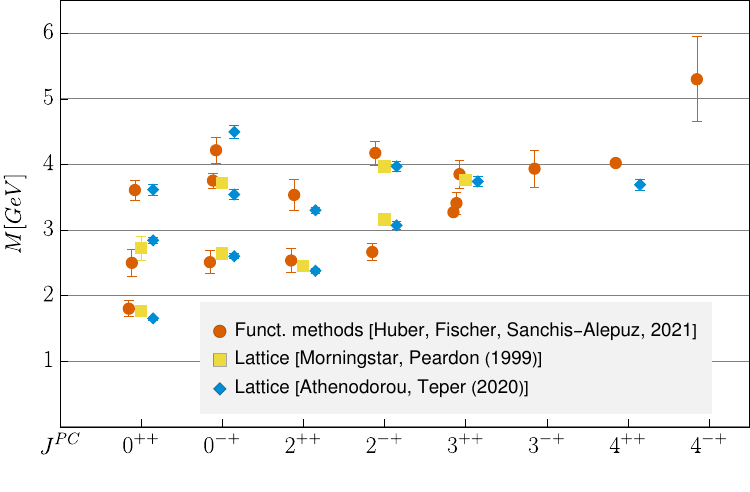}
	\caption{Left: Ghost and gluon dressing functions \cite{Huber:2020keu,Sternbeck:2006cg}.
	Right: Glueball spectrum of pure Yang-Mills theory \cite{Morningstar:1999rf,Athenodorou:2020ani,Huber:2021yfy}.
	All results have been rescaled to the same value of the Sommer scale $r_0=1/(418\,\text{MeV})$.
	}
	\label{fig:props_spectrum_YM}
\end{figure}

\begin{figure}[t]
	\centering
	\includegraphics[width=0.48\textwidth]{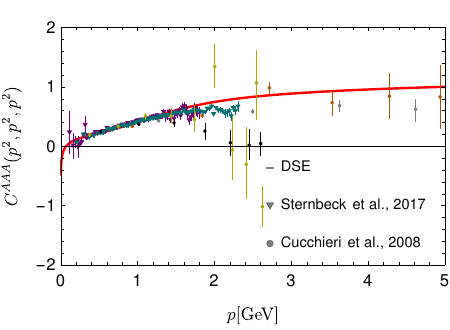}\hfill
	\includegraphics[width=0.48\textwidth]{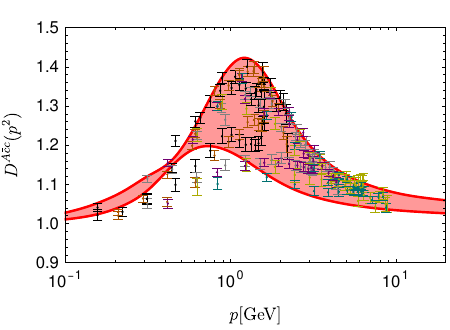}
	\caption{
	Left: Three-gluon vertex at the symmetric point \cite{Cucchieri:2008qm,Sternbeck:2017ntv,Huber:2020keu}.
	Right: Ghost-gluon vertex with full kinematic dependence \cite{Maas:2019ggf,Huber:2020keu}.
	}
	\label{fig:tg_ghg_YM}
\end{figure}

The system of correlation functions depicted in \fref{fig:3PI_3l_YM_EOM} was solved in \cite{Huber:2020keu}.
Fig.~\ref{fig:props_spectrum_YM} shows the propagator dressing functions and \fref{fig:tg_ghg_YM} the dressing functions of the three-gluon and ghost-gluon vertices.
The vertices were calculated using kinematic variables based on the $S_3$ permutation group \cite{Eichmann:2014xya}.
The kinematic dependence on the doublet variables is represented by the bands in the figures.
The three-gluon vertex exhibits a particularly small dependence on the doublet variables \cite{Eichmann:2014xya,Huber:2020keu,Aguilar:2023qqd}.
On the contrary, for the ghost-gluon vertex this dependence is quite pronounced \cite{Huber:2012kd,Aguilar:2018csq}.
By now, also lattice results for the four-gluon vertex exist which compare favorably to the functional ones \cite{Huber:2025kwy}.
In general, the results do not only agree well with results obtained with lattice calculations, as demonstrated in \fref{fig:props_spectrum_YM} and \fref{fig:tg_ghg_YM}, but also with results from the complementary functional system of equations obtained from the FRG \cite{Cyrol:2016tym}.
In addition, there is considerable evidence of their stability against further extensions of the truncation.
Both these aspects are discussed in detail in \cite{Huber:2025kwy}.

Pure Yang-Mills theory has only one scale, which is generated by dimensional transmutation.
The complete calculation can thus be performed with arbitrary units which are then fixed a posteriori.
For the purpose of comparison between different methods, it is necessary to use the same units.
It is advantageous to take over the units of lattice calculations, typically fixed via the string tension, by matching the positions of the peaks of the gluon dressing functions.
This method relies on the fact that the results of the two methods agree well.
In the plots, this fixes then the horizontal axis.
On the vertical axis, the results can be shifted in the comparison plots as this corresponds to a change in the renormalization condition.

\subsection{Glueball spectrum}

With the results from the system of equations in \fref{fig:3PI_3l_YM_EOM} as input one can then solve the corresponding BSEs with the kernels depicted in \fref{fig:kernels_YM_3PI}.
As discussed in Sec.~\ref{sec:solveBSEs}, one solves an eigenvalue problem as a function of $P^2$ where $P$ is the total glueball momentum.
However, with the current input, this is restricted to $P^2>0$ since the correlation functions are only available for space-like momenta.
An extension to complex momenta, see Sec.~\ref{sec:complexMomenta}, is only available for simpler truncations \cite{Fischer:2020xnb,Horak:2022myj}.
To overcome this, an extrapolation of the eigenvalue curve to the physical values $P^2=-M^2<0$ is used.
The reliability of such an extrapolation was successfully demonstrated for a meson in a truncation where this problem is absent \cite{Huber:2020ngt}.
The error given for the glueball masses stems from the extrapolation.

Originally, the BSEs were solved only with the one-particle exchange diagrams of the kernels \cite{Huber:2020ngt,Huber:2021yfy}.
The inclusion of more diagrams for the three lightest glueballs was subsequently performed in \cite{Huber:2025kwy}.
The tiny changes of less than 2\% that were found for the masses are a nice confirmation of the stability of the results and illustrate apparent convergence at work.
The final results for the spectrum of glueballs with $J^\mathsf{CP}=J^{+\pm}$, $J=0,2,3,4$ are shown in \fref{fig:props_spectrum_YM}.
In other settings, one concludes from the Landau-Yang theorem \cite{Landau:1948kw,Yang:1950rg} that a three-gluon bound state equation is necessary to describe $J=1$.
However, although it does not apply here as the gluons are not on-shell, no solutions for $J=1$ were found.
A three-gluon equation would thus be interesting to investigate, which also describes states with negative charge parity \cite{Huber:2020ngt}.
The obtained spectrum is in good agreement with corresponding lattice results also shown in \fref{fig:props_spectrum_YM}.

\section{Conclusions}
\label{sec:conclusions}

In this article, I have discussed the use of functional equations in particle physics.
I introduced Dyson-Schwinger equations, $n$PI effective actions, bound state equations, and the functional renormalization group.
Due to the extensiveness and complexity of these topics, I could only scratch the surface and singled out a few aspects of particular interest for beginners in the field.
Among them are overviews over truncation schemes and some computational techniques.
The workflow of functional methods was exemplified with the glueball spectrum of pure Yang-Mills theory.
The truncation scheme, the calculation of the required correlation functions and the solution of the bound state equations have been discussed.
The good agreement with corresponding lattice results illustrates the potential of functional equations as a first-principles method.

What makes functional methods particularly attractive is their complementarity to lattice calculations.
They are continuum methods and thus lack the complications typical for lattice calculations related to discretization.
Quark masses can be varied continuously and easily and at nonzero chemical quark potential there is no sign problem.
Beyond that, one should also keep in mind that computational resources are in many cases much lower than for lattice calculations.
However, functional methods come of course with their own challenges, the foremost being the need to truncate the equations.
In this regard, the notion of apparent convergence has been introduced for which already a few examples exist and work is going on to extend setups for more systems.
Even when not in the ideal situation of having a fully self-contained system of equations that proves to be stable against further extensions, calculations using models or input from other methods can provide not only useful insights into concepts and mechanisms but also access to comprehensive quantitative predictions of, for example, hadron masses or their structure, see, e.g., \cite{Eichmann:2016yit,Eichmann:2025wgs}.

Although explicit examples in this article have been taken from hadron physics, this should not distract from the fact that functional equations are useful tools in other areas of particle physics and beyond as well.
QED was studied in three and four dimensions, for instance, in \cite{Maris:1995ns,Bonnet:2011hh,Kizilersu:2014ela,Jia:2017niz}.
Questions like the triviality problem of the Higgs sector in the Standard Model or beyond the Standard Model scenarios were investigated, see, e.g., \cite{Hopfer:2014zna,Goertz:2024dnz}.
The phase structure of QCD is also a heavily studied topic where functional methods can profit from their complementary nature to lattice calculations, especially the abovementioned absence of a sign problem, see, e.g., \cite{Fischer:2018sdj,Dupuis:2020fhh,Fischer:2026uni}.
For the unification of gravity and quantum physics asymptotic safety is a widely studied scenario, see, e.g., \cite{Eichhorn:2018yfc} for an overview.
Further topics include condensed matter or cold atoms, see the review \cite{Dupuis:2020fhh} for an overview, or even chemistry \cite{Blase:2020ptc}.

The steady progress that has been made with functional methods over the last three decades builds a solid foundation for tackling many problems of particle physics and beyond.
A significant step forward is the extension to fully self-contained calculations, while phenomenologically driven model-based ones continue to provide relevant insights.
More research will be done to extend advances in one area to others, from vacuum to nonzero temperature and density, from space-like to time-like, from mesons and baryons to many-quark states, from static to dynamical observables.

\begin{ack}[Acknowledgments]
It is a pleasure to thank all my collaborators and colleagues who have contributed to my understanding of the topics discussed here, in particular Reinhard Alkofer, Jens Braun, Davide Campagnari, Anton K. Cyrol, Gernot Eichmann, Christian S. Fischer, Axel Maas, Mario Mitter, Jan M. Pawlowski, Hugo Reinhardt, H\`elios Sanchis Alepuz, Bernd-Jochen Schaefer, Kai Schwenzer, Lorenz von Smekal, and Richard Williams.
I am grateful to Jens Braun and Bernd-Jochen Schaefer for a critical reading of some parts of the manuscript and helpful comments.
This work was supported in part by the Deutsche Forschungsgemeinschaft (DFG, German Research Foundation) under Contract No. HU 2176/3-1.
Feynman diagrams were made with \textit{JaxoDraw} \cite{Binosi:2003yf}.
\end{ack}

\bibliographystyle{utphys_mod}
\bibliography{literature_funMethodsBeginner}

\end{document}